\documentclass[preprint,showpacs,amsmath,amssymb,aps,floats,floatfix,nofootinbib]{revtex4-1}
\usepackage{amsmath,amsfonts,amssymb,bm}
\usepackage{graphicx}
\usepackage{epstopdf}
\usepackage{epsfig}
\usepackage{color}
\usepackage[dvipsnames,svgnames,x11names]{xcolor}
\usepackage{slashed}
\usepackage{url}
\usepackage{subfigure}
\usepackage{cleveref}
\usepackage{multirow}
\usepackage[normalem]{ulem}

\graphicspath{{figs/}}  

\def\cm{\,\mathrm{cm}}
\def\km{\,\mathrm{km}}
\def\kpc{\,\mathrm{kpc}}
\def\sec{\,\mathrm{s}}

\def\TeV{\,\mathrm{TeV}}
\def\GeV{\,\mathrm{GeV}}
\def\GV{\,\mathrm{GV}}

\makeatother

\allowdisplaybreaks

\def\planss{Planet. Space Sci.\ }

\def\apj{Astrophys.\ J.\ }

\begin{document}

\title{Investigating the dark matter signal in the cosmic ray antiproton flux with the machine learning method} \author{Su-Jie Lin$^{1,2}$, Xiao-Jun Bi$^{1,2}$, Peng-Fei Yin$^1$}
\affiliation{$^1$Key Laboratory of Particle Astrophysics, Institute of High Energy Physics, Chinese Academy of Sciences, Beijing 100049, China}
\affiliation{$^2$School of Physical Sciences, University of Chinese Academy of Sciences, Beijing 100049, China}

\begin{abstract}
We investigate the implications on the dark matter (DM) signal from the AMS-02 cosmic antiproton flux.
Global fits to the data are performed under different propagation and hadronic interaction models.
The uncertainties from the injection spectrum, propagation effects and solar modulation of the cosmic rays are taken into account comprehensively.
Since we need to investigate extended parameter regions with multiple free parameters in the fit, the machine learning method is adopted to maintain a realistic time cost.
We find all the effects considered in the fitting process interplay with each other, among which the hadronic interaction model is the most important factor affecting the result.
In most hadronic interaction and CR propagation models no DM signal is found with significance larger than $2\sigma$ except that the EPOS-LHC interaction model requires a more than $3\sigma$ DM signal with DM mass around $1\TeV$.
For the diffusive reacceleration propagation model there is a highly significant DM signal with mass around $100\GeV$.
However, the signal becomes less than $1\sigma$ if we take a charge dependent solar modulation potential in the analysis.
\end{abstract}


\maketitle

\section{Introduction} \label{section:Introduction}

Cosmic ray (CR) antiproton is a very promising probe in the dark matter (DM) indirect detection.
In recent years, a great progress of the measurement has been made by the AMS-02 experiment, which precisely measured the CR $\bar{p}/p$ ratio and $\bar{p}$ flux up to $\sim 450$~GeV.
Based on these results, lots of theoretical studies have been performed to investigate the possible contribution from DM in the antiproton flux~\cite{Giesen:2015ufa,Jin:2015mka,Chen:2015cqa,Ibe:2015tma,Hamaguchi:2015wga,Kohri:2015mga,Kappl:2015bqa,Lu:2015pta,Lin:2015taa,Cui:2016ppb,Cuoco:2016eej,Feng:2016loc,Huang:2016tfo,Lin:2016ezz,Cui:2018klo,Cui:2018nlm,Cuoco:2019kuu,Cholis:2019ejx}.
As the deviation between the expected secondary CR antiproton flux produced by the astrophysical processes and the data is not significant, any slight change of the background expectation could influence the DM implication.
Therefore, the theoretical uncertainties of the secondary CR antiproton flux, mainly arising from the CR propagation process and the hadronic interactions between the CR particles and interstellar medium (ISM), are very important in the determination of the DM contribution.

Although the principle of the strong interaction has been well described by the quantum chromodynamics (QCD), only the processes with large momentum transfers can be perturbatively calculated.
The production processes of secondary CR antiprotons, which involve the forward scattering processes with multi-particle production, are always calculated under extra simplified assumptions or using some empirical parametrizations.
This kind of processes has been a long time difficulty in the cosmic ray study.
Many phenomenological models with different simplified assumptions have been constructed in the literature~\cite{Kalmykov:1997te,Ostapchenko:2004ss,Engel:1992vf,Engel:1994vs,Bopp:2005cr,Werner:2005jf,Pierog:2013ria}.
In addition, several empirical parametrizations only aiming to reproduce the antiproton production cross section have also been developed~\cite{1983JPhG....9.1289T,diMauro:2014zea,Winkler:2017xor}.
On the other hand, the CR propagation can be determined by the fit to the observed secondary-to-primary nuclei ratios like B/C and (Sc+Ti+V)/Fe, and unstable-to-stable ratios of secondary nuclei like $^{10}\mathrm{Be}/^9\mathrm{Be}$ and $^{26}\mathrm{Al}/^{27}\mathrm{Al}$~\cite{Strong:1998pw,1990cup..book.....G,DiBernardo:2009ku,Maurin:2001sj}.
With the constraints of current data, there remains a considerable acceptable region for the varying propagation parameters~\cite{Yuan:2017ozr}.

With these uncertainties involved, we have found that several hadronic models can well fit the AMS-02 antiproton data, and derived the upper-limits on the DM annihilation cross section in a previous work~\cite{Lin:2016ezz}.
However, in the scenarios with the other hadronic models or astrophysical configurations, the secondary CR antiproton could be insufficient in some energy ranges.
In these cases, it is interesting to explore the favored region in the DM parameter space by investigating how an extra DM contribution could improve the fit~\cite{Cui:2016ppb,Cuoco:2016eej,Cui:2018klo,Cuoco:2019kuu,Cholis:2019ejx}.

To set the upper-limits, we could just separately estimate the antiproton fluxes with thousands of benchmark astrophysical parameter sets within the 2$\sigma$ confidence region accounting for the B/C fit, and use the band spanned by the corresponding limits to indicate the astrophysical uncertainties.
However, claiming an excess and investigating a favored DM parameter region should be prudent and much more complicated.
As the favored region is sensitive to the secondary CR antiproton spectrum, investigating thousands of benchmark astrophysical parameter sets is far from enough.
It is necessary to vary the astrophysical parameters when searching for the favored region in order to reach a comprehensive solution~\cite{Cui:2016ppb,Cuoco:2016eej,Cui:2018nlm,Cuoco:2019kuu}.

In this work, we try to investigate this favored region by performing a likelihood optimization in the $m_\chi-\langle\sigma v\rangle$ plane, in which the profile likelihood for each given $(m_\chi, \langle\sigma v\rangle)$ is obtained by varying the astrophysical parameters about ten thousand times.
Since the CR propagation equation solution for each set of astrophysical parameter always costs about several minutes by using the well-developed numerical packages like GALPROP~\cite{Strong:1998pw,Moskalenko:1997gh} or DRAGON~\cite{Evoli:2016xgn}, it is very time-consuming to perform the whole scanning analysis.
In order to solve this efficiency problem, we adopt the machine learning method in the analysis, which works through a training and predicting process.
We perform the CR propagation calculation using GALPROP for about ten thousand sets of the astrophysical parameters, and take these results to train the machine learning model.
Then we utilize the well-trained model to predict the antiproton flux for given astrophysical parameters in the scanning analysis.
Since the calculation of the machine learning model is much faster than that of GALPROP, this method is able to significantly improve the efficiency.
With this technique introduced, it then is possible to statistically study the DM implication in the AMS-02 antiproton data for different hadronic and CR propagation models.

This paper is organized as follows.
In Sec.~\ref{section:cosmic_ray_antiproton_flux}, we describe the physical processes related to the CR antiproton flux, including the CR propagation and hadronic models, and DM annihilation.
In Sec.~\ref{section:machine_learning}, we introduce the adopted machine learning method in brief, and show the training quality of the machine learning model.
In Sec.~\ref{section:implications_for_dm_annihilation}, we show the final scan results for different hadronic and CR propagation models, and discuss the implications of DM contribution in the AMS-02 antiproton flux.
Finally, we give the summary in Sec.~\ref{section:conclusions}.

\section{Cosmic ray antiproton flux} \label{section:cosmic_ray_antiproton_flux}

The CR antiprotons are considered to originate from the hadronic interaction between primary CR particles and ISM, perhaps along with the additional contribution from DM annihilation/decay.
These antiprotons would undergo the processes like propagation and solar modulation before arriving at the Earth.

In this section, we would introduce the CR propagation, solar modulation, hadronic interaction, and DM annihilation processes considered in this work in detail.
The relevant adopted parameters of the CR propagation and solar modulation processes are based on our previous work~\cite{Yuan:2017ozr}.

\subsection{Propagation of Galactic cosmic rays} \label{subsection_propagation_of_galactic_cosmic_rays}
The Galactic CRs are believed to originate from the supernova remnants (SNRs), and propagate diffusively in the Galactic magnetic field (GMF) that is extended within a cylindrical propagation halo.
Such a diffusive propagation together with the accompanying effects can be described with the equation~\cite{Strong:2007nh}
\begin{eqnarray}
\frac{\partial \psi}{\partial t} &=& Q(\mathbf{x}, p) + \nabla \cdot ( D_{xx}\nabla\psi - \mathbf{V}_{c}\psi )
+ \frac{\partial}{\partial p}\left[p^2D_{pp}\frac{\partial}{\partial p}\left(\frac{\psi}{p^2}\right)\right]
\nonumber\\
&& - \frac{\partial}{\partial p}\left[ \dot{p}\psi - \frac{p}{3}(\nabla\cdot\mathbf{V}_c)\psi \right]
- \frac{\psi}{\tau_f} - \frac{\psi}{\tau_r},
\label{propagation_equation}
\end{eqnarray}
where $\psi$ is the CR density per momentum interval, $Q(\mathbf{x}, p)$ is the CR source term which is assumed to follow a three-pieces broken power-law with respect to the rigidity, the $D_{xx}$ term describes the diffusion effect, the $\mathbf{V}_c$ terms describe the convection effect, the $D_{pp}$ term describes the reacceleration effect, the term with $\dot{p}\equiv \mathrm{d}p/\mathrm{d}t$ is the momentum loss term, and the time scales $\tau_f$ and $\tau_r$ characterize fragmentation processes and radioactive decays, respectively.

This equation applies to all kinds of CR particles.
As the source term for the secondary particles depends on the fragmentation and decay of the primary particles, the secondary fluxes linearly depend on the primary fluxes.
Therefore, their ratios could be used to characterize the propagation process.
In practice, the secondary-to-primary ratios like B/C and the unstable-to-stable ratios like $^{10}\mathrm{Be}/^9\mathrm{Be}$ are usually adopted to constraint the coefficients included in Eq.~\ref{propagation_equation}~\cite{Putze:2010zn,Trotta:2010mx}.
In the previous work~\cite{Yuan:2017ozr}, we have fitted the B/C ratio observed by AMS-02~\cite{Aguilar:2016vqr} with the different propagation models, and found that the diffusion plus reacceleration (DR) model works better than the diffusion plus convection (DC) model.
Therefore, we only consider the reacceleration effect in this work.

The spatial distribution of the primary CR sources $Q(\mathbf{x}, p)$ is supposed to follow the SNR distribution
\begin{equation}
f(r,z) = \left( \frac{r}{r_\odot} \right)^a \exp\left[ -\frac{b(r-r_\odot)}{r_\odot} \right]\exp\left( -\frac{\left|z\right|}{z_s} \right),
\label{spatial_distribution}
\end{equation}
where $r_\odot=8.5\kpc$ and $z_s\simeq 0.2\kpc$ are the distance from the Sun to the Galactic Center and the characteristic height of the Galactic disk, respectively.
In this work, we set $a=1.25$ and $b=3.56$ based on the Fermi-LAT gamma-ray data following the Ref.~\cite{Trotta:2010mx}.

The diffusion coefficient $D_{xx}$ of the CR particles is empirically parameterized as a power-law function of rigidity $R$ along with a low energy modification~\cite{Maurin:2010zp}
\begin{equation}
D_{xx} = D_0\beta^\eta \left( R/R_0 \right)^{\delta}.
\end{equation}
The power-index $\delta$ is expected to be $1/3$ for a Kolmogorov spectrum of interstellar turbulence, or $1/2$ for a Kraichnan cascade.
However, we just treat it as a free parameter and determine it by the observed B/C ratio.
$\beta$ is the CR particle velocity in units of the light speed that would tends to 1 in the relativistic limit.
The factor $\beta^\eta$ is introduced to describe the effect of turbulence dissipation, where $\eta$ is assumed to be 1 by default.
But it is found that changing the value of $\eta$ could help to improve the fitting of secondary CR particles~\cite{DiBernardo:2010is}.
Therefore, in this work we also discuss the propagation model with a free $\eta$, which is referred to as the DR-2 model following the Ref.~\cite{Yuan:2017ozr}.

The CR reacceleration effect is a second-order Fermi acceleration effect caused by the collisions between CR particles and the interstellar random weak hydrodynamic waves.
It is described by a diffusion in the momentum space whose diffusion coefficient $D_{pp}$ is related to $D_{xx}$ by~\cite{1994ApJ...431..705S}
\begin{equation}
D_{pp} D_{xx}=\frac{4p^2v^2_{A}}{3\delta(4-\delta^2)(4-\delta)\omega }\\,
\label{reacceleration}
\end{equation}
where $v_{A}$ is the Alfv\'{e}n velocity, and $\omega$ is the energy density ratio between the magneto-hydrodynamic wave and the magnetic field. In practical calculation, $\omega$ could be absorbed into the $v_{A}$.

In all, the relevant propagation parameters include the normalizing factor $D_0$, power-index $\delta$, reference rigidity $R_0$ for the diffusion coefficient, the Alfv\'{e}n velocity $v_A$, the half height of propagation halo $z_h$, and perhaps the parameter $\eta$.
In this work, we adopt the DR and DR-2 models from Ref.~\cite{Yuan:2017ozr}, where the reference rigidity $R_0$ is taken to be $4\GV$.
The corresponding best fitting parameters are listed in Table.~\ref{tab:propagation_parameters}.

\begin{table*}
  \centering
  \begin{tabular}{cccccccc}
    \hline
         & $D_0$                   & $\delta$ & $z_h$  &    $v_A$       & $\eta$ & & $\phi_+$ \\
         \cline{2-6}\cline{8-8}
         & $10^{28}\cm^2\sec^{-1}$ &          & $\kpc$ & $\km\sec^{-1}$ &        & &  $\GV$ \\
    \hline
    DR   &     7.24                & 0.38     & 5.93   &    38.5        &    1   & &  0.527 \\
    DR-2 &     4.16                & 0.5      & 5.02   &    18.4        &  -1.28 & &  0.636 \\
    \hline
  \end{tabular}
  \caption{The best fits for the propagation parameters and the positive modulation potential $\phi_+$ in the DR and DR-2 models, obtained from the Ref.~\cite{Yuan:2017ozr}.}
  \label{tab:propagation_parameters}
\end{table*}

\subsection{Solar modulation} \label{section:solar_modulation}

The CR particles would also be affected by the heliospheric magnetic field (HMF) when they enter the heliosphere.
Such an effect that mainly influences the particles with rigidity $R\lesssim 20\GV$ is called solar modulation, and can be described by the Parker's transport equation~\cite{1965P&SS...13....9P}.
This equation was firstly solved by Gleeson et. al.~\cite{Gleeson:1968zza} using the force field approximation (FFA), in which the solar modulation potential $\phi$ is the sole parameter.

However, this solution is derived under the spherical symmetric assumption, and has neglected the drift effect caused by the configuration of HMF.
A recent study, which solved the Parker's transport equation with the realistic simulation~\cite{Potgieter:2014pka}, has found that this drift effect would indeed lead to a charge-sign dependent behaviour in the CR spectra.
Therefore, adopting the FFA with only one modulation potential $\phi$ is insufficient to describe all the CR particles.
In this work, we adopt the FFA to deal with the solar modulation, but assume two modulation potentials $\phi_+$ and $\phi_-$ for the positive and negative charged particles respectively.

The strength of solar modulation would vary with the solar activity, thus the value of modulation potential depends on the observation period of the experiment.
In Ref.~\cite{Yuan:2017ozr}, we assumed that the modulation potential $\phi$ is linearly correlated with the sunspots' number $N$, such that $\phi = \phi_0 + \phi_1\cdot N / N_{\mathrm{max}}$, and fitted the data from different periods for the coefficients $\phi_0$, $\phi_1$.
For the period of AMS-02 $\bar{p}/p$ observation, the corresponding potentials are $\sim0.527\GV$ and $0.636\GV$ for the DR and DR-2 models, respectively, as shown in Table.~\ref{tab:propagation_parameters}.
As these values are derived by fitting the proton and B/C data, they are in fact the modulation potential for positive particles $\phi_+$.
The negative one $\phi_-$ is set to be a free parameter vary from $0.5\phi_+$ to $1.5\phi_+$ in our analysis.

\subsection{Hadronic models}

The hadronic interaction processes between the primary CR particles and the environment gas could not be directly calculated in the first principle with QCD.
Basing on collider data, many phenomenological and empirical models for these processes have been developed to deal with this problem.

The phenomenological models describe the fragmentation process of nucleon based on quantum field theory along with some effective assumption.
These models are derived for different purposes.
HERWIG\cite{Corcella:2000bw}, PYTHIA\cite{Sjostrand:2006za} and SHERPA\cite{Gleisberg:2008ta} focus on the hard-scattering processes for high energy physics studies.
QGSJET01\cite{Kalmykov:1997te}, QGSJET II\cite{Ostapchenko:2004ss,Ostapchenko:2005nj} and SIBYLL\cite{Engel:1992vf,Fletcher:1994bd,Ahn:2009wx} focus on the bulk production of soft particles for simulating extensive air showers caused by high energy CR particles.
PHOJET\cite{Engel:1994vs,Engel:1995yda,Engel:1995sb}, DPMJET\cite{Bopp:2005cr} and EPOS\cite{Werner:2005jf,Pierog:2013ria} fall in between the two categories.

As we aim to calculate the secondary CR particles which are dominantly contributed by the soft production, the models focusing on the hard-scattering processes are not taken into account.
The efficiency of the other models in the concerned energy region could be checked by comparing them with the collider data at  $\sqrt{s}\sim\mathcal{O}(10\GeV)$.
These data are only available at several benchmark colliding energies in the experiments.
The CERN Intersecting Storage Rings (ISR) experiment has measured the $pp\rightarrow\bar{p}X$ cross section at $\sqrt{s}=53\GeV$ in mid-1970s~\cite{Albrow:1973kj}.
In recent years, the fixed-target experiment NA49 at CERN Super Proton Synchrotron (SPS) has also measured such a cross section at beam momentum of $158\GeV/c$, equaling to $\sqrt{s}=17.27\GeV$ in the center-of-mass system~\cite{Anticic:2009wd}.

In the previous work, we have taken the EPOS, QGSJET and SIBYLL into account and discussed their efficiencies~\cite{Lin:2016ezz}.
We approached these models through the common interface provided by the CRMC package~\cite{CRMCWeb}.
Both the two versions of EPOS, namely EPOS 1.99~\cite{Pierog:2009zt} and EPOS LHC~\cite{Pierog:2013ria}, have been tuned to fit the results from the fixed-target experiments~\cite{Anticic:2009wd,Laszlo:2007ib}, and could reasonably reproduce all the measured antiproton cross sections mentioned above.
The QGSJETII has only considered the antiproton cross sections in a modified version, namely QGSJETII-04m~\cite{Kachelriess:2015wpa}.
This modified model fits the low energy antiproton cross section measurements well, but would always tend to overestimate the low energy secondary CR antiproton flux.
The SIBYLL can also fit the antiproton cross sections well, but cannot extend to the extremely low energy region with a beam momentum $\lesssim 60\GeV/c$.
In all, the only two phenomenological models discussed in this work are chosen to be EPOS 1.99 and EPOS LHC.

Another approach to describe the antiproton production cross sections is the empirical model that fits the measurements with certain analytical parametrization~\cite{1983JPhG....9.1289T,Duperray:2003bd,diMauro:2014zea,Winkler:2017xor}.
In the following, we select several empirical models as the typical cases,
including the default hadronic model embedded in GALPROP by Tan \& Ng~\cite{1983JPhG....9.1289T},
the model fitting to the latest NA49~\cite{Anticic:2009wd} and BRAHMS~\cite{Arsene:2007jd} data by di Mauro et al.~\cite{diMauro:2014zea},
and the model carefully considering the effects of anti-hyperons and isospin by Winkler~\cite{Winkler:2017xor}.
There are in fact two similar parametrizations in Ref.~\cite{diMauro:2014zea}, but we just choose the expression in its Eq.~13 here.

These empirical models only describe the antiproton productions from $pp$ collisions.
In order to estimate the secondary CR antiprotons, the contributions of Helium which occupy about 10\% of both the CR and the ISM should be considered.
Following the procedure of GALPROP~\cite{GALPROPTheoryWebSite}, we adopt the code CROSEC by Barashenkov \& Polanski to derive the antiproton production cross section of $p$-He and He-He collisions from that of $pp$ collisions.

Although empirical models could always reproduce the accelerator data better compared with phenomenological models at the available collision energies,
phenomenological models that constructed with certain physical principles could guarantee a more reasonable estimation when extended to the unexplored collision energy regions.
Thus we involve both the two kinds of models in this analysis, including EPOS 1.99, EPOS LHC, and the empirical models by Tan \& Ng, di Mauro et al., and Winker.

\subsection{Dark matter annihilation}

The DM particles could directly annihilate into standard model particles, and produce stable final states like protons/antiprotons, electrons/positrons and photons through the hadronization and decay.
Since all the DM annihilation channels to light quarks would lead to similar antiproton spectra, we choose the $b\bar{b}$ channel as a benchmark case in this work.
The initial injected antiproton spectra are adopted from PPPC 4 DM ID~\cite{Cirelli:2010xx}.

The contribution from DM annihilation also depends on the density profile of the Galactic DM halo.
Here we adopt the Navarro-Frenk-White (NFW) profile~\cite{Navarro:1996gj}
\begin{equation}
  \rho(r)=\frac{\rho_s}{(r/r_s)(1 + r/r_s)^2}
  \label{eq:NFW}
\end{equation}
with $r_s=20\kpc$ and $\rho_s=0.35\GeV$.
This profile corresponds to a local DM density of $0.4\GeV\cm^{-3}$ near the Solar system, which is consistent with the recent constraints from the Galactic rotation curve~\cite{Nesti:2013uwa,Benito:2019ngh,Karukes:2019jxv}.

\section{Machine learning} \label{section:machine_learning}

In this work, we attempt to investigate the DM implication through a profile likelihood scan in the $m_\chi-\langle\sigma v\rangle$ plane.
The profile likelihood for each $(m_\chi, \langle\sigma v\rangle)$ used in the scan is obtained by marginalizing over the propagation parameters near the best fit shown in Table.~\ref{tab:propagation_parameters}, which would require us to solve the propagation equation $\sim10^4$ times.
Using the numerical package GALPROP~\cite{Strong:1998pw}, each propagation solution would cost about one or several minutes.
Such a cost is unacceptable for the parameter scan mentioned above, thus a more efficient method is required.
We adopt the machine learning method for this purpose.

The supervised machine learning is always used to perform general classifications or regressions.
Such regressions could be applied to predict the wanted CR fluxes in order to release us from the time-consuming numerical solutions.
We have tested several widely used learning algorithms for this problem,
including the support-vector machine (SVM)~\cite{Cortes1995}, K-nearest neighbors (KNN)~\cite{Altman:1992ait}, decision tree (DT)~\cite{Kamiński2018}, random forest~\cite{Ho:1998trs}, and artificial neural networks (ANN)~\cite{McCulloch1943}.
We find that, focusing on the 2$\sigma$ confidence region favored by the B/C data, the relatively simple SVM algorithm works the best.
The reason is that this interested problem is almost a linear regression problem, as the solutions of Eq.~\ref{propagation_equation} could be perturbatively expanded within the narrow region near the best-fit propagation parameters.
Therefore, in our scanning analysis, we choose the support-vector regression (SVR) algorithm that based on the SVM.
The reader is referred to the Appendix~\ref{section:SVM} for the detailed description of this algorithm.
We use the package LIBSVM~\cite{Chang:2011:LLS:1961189.1961199} to approach training and predicting process of the SVR.

Note that the SVR algorithm can be only applied to the single target problem.
However, the propagation solution is described as a multi target problem, whose targets are the CR fluxes at different energies.
In order to involve the relation between different targets, we adopt the stacked single-target (SST) algorithm in the calculation~\cite{Spyromitros-Xioufis2016}.
In the SST procedure, we train each target separately, and then combine the predictions of these targets together with the original input to perform a second training.
This second training would correct the prediction of each single target with the information of the other targets.

\subsection{2$\sigma$ confidence region for the astrophysical parameters}
In this work, our scans are constrained within the 2$\sigma$ confidence region derived from the fit to the $\mathrm{B}/\mathrm{C}$, $\mathrm{Be}^{10}/\mathrm{Be}^9$, and proton flux observation~\cite{Yuan:2017ozr}.
For each propagation model, we adopt a set of samples generated by the Monte Carlo Markov-Chain (MCMC) algorithm from our previous work Ref.~\cite{Yuan:2017ozr}.
In each sample, there are a set of propagation, injection, and solar modulation parameters together with the corresponding $\chi^2$ for the $\mathrm{B}/\mathrm{C}$ fitting.
The best $95\%$ of the samples indicate the 2$\sigma$ confidence region.

In the implementation, such a 2$\sigma$ confidence region is also recognized with machine learning.
We perform a training on $\bm{\theta}_i\mapsto\chi^2_i$ with the given samples,
and then use the predicting function $\hat{h}_{\chi^2}(\bm{\theta})$ to check whether the point $\bm{\theta}$ falls inside the region.

Note that most of the MCMC samples would concentrate around the best fit parameter point and is sparse in the other region.
We thus append a set of evenly distributed samples in order to obtain a better prediction in the whole region.
In all, there are $20,000\sim30,000$ MCMC samples and 5000 appended samples for each propagation model.
Such combined sample sets would be used as the training sets for all the following analyses.

\subsection{Training for the background spectrum} \label{section:training_for_the_background_spectrum}

\begin{figure}[!htbp]
  \centering
  \includegraphics[width=0.45\textwidth]{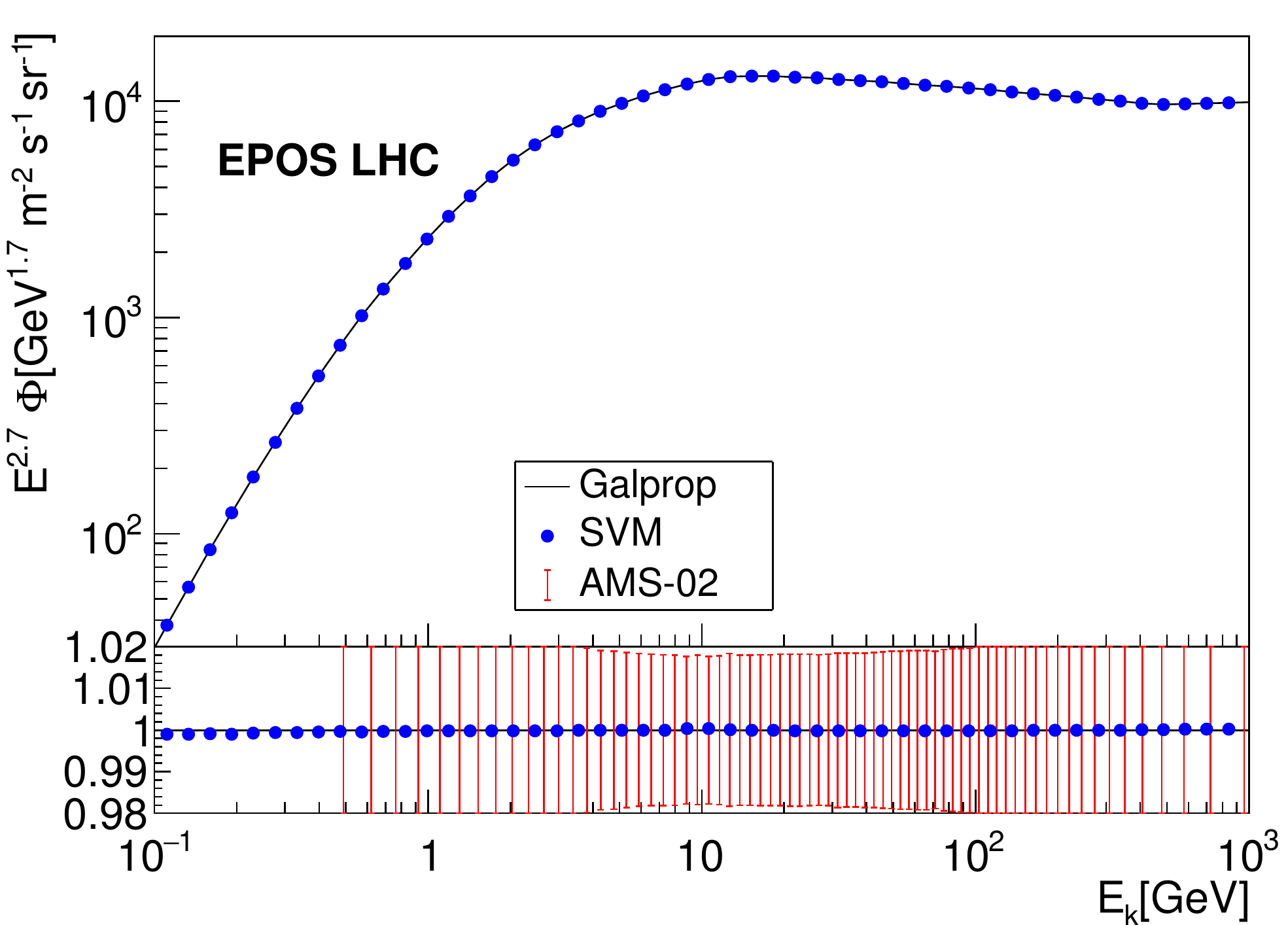}
  \includegraphics[width=0.45\textwidth]{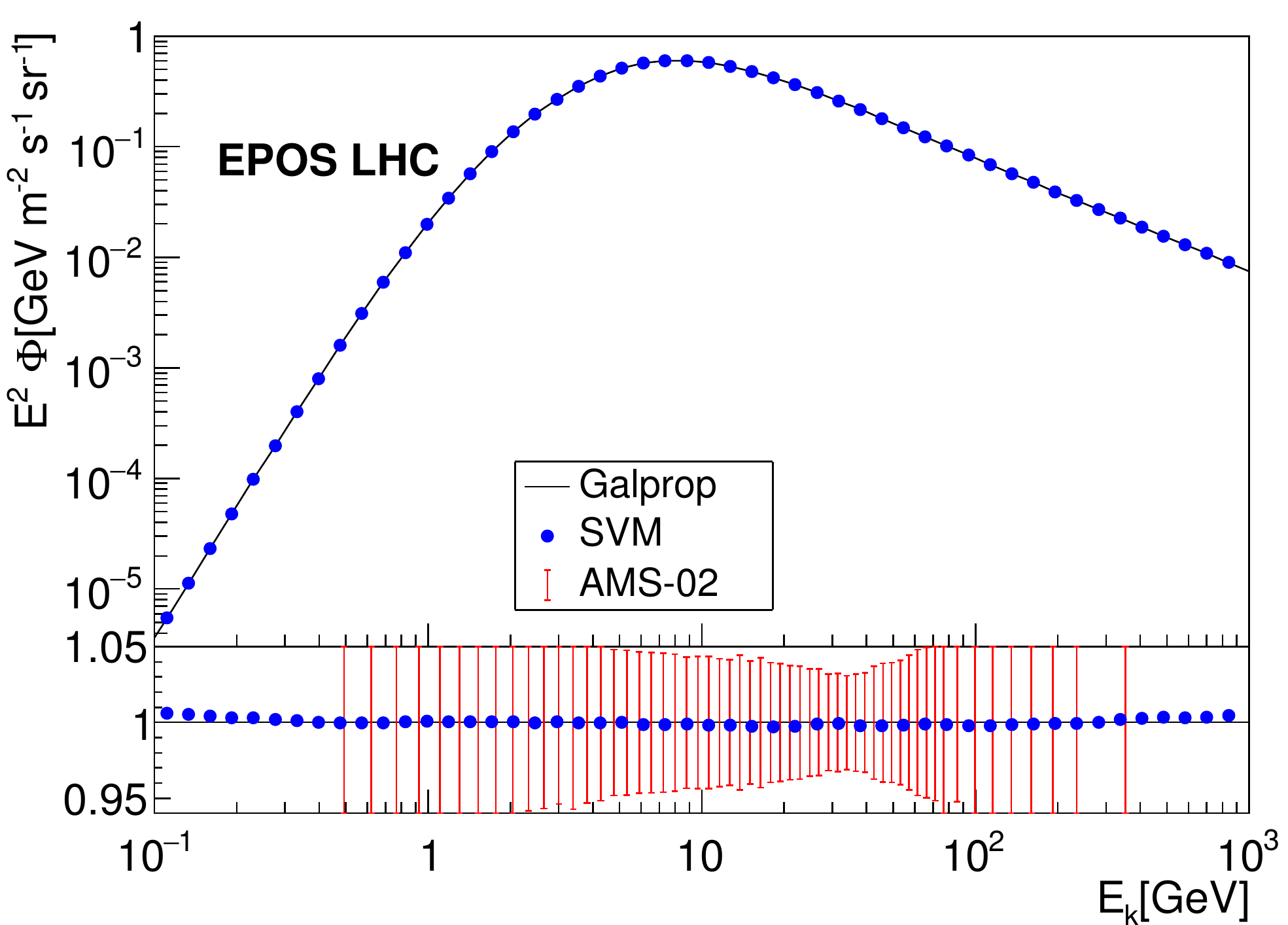}
  \caption{An example of the CR proton (left panel) and antiproton (right panel) fluxes predicted by the well-trained SVM models, in comparison with the GALPROP results and the corresponding uncertainty of the AMS-02 measurement~\cite{Aguilar:2015ooa,Aguilar:2016kjl}. The propagation parameter set is randomly specified in DR-2 propagation model. The hadronic model is taken to be the EPOS LHC model.}
  \label{fig:predict_example}
\end{figure}

Selecting a sequence of energies which cover the whole energy range of AMS-02, we take the corresponding proton, secondary antiproton and tertiary antiproton fluxes as the training targets.
In order to keep the problem simple, we involve no solar modulation effect in the training.
As mentioned above, we adopt the SST procedure in which the neighbor fluxes of each point are involved in the training.
In Fig.~\ref{fig:predict_example}, we show a random prediction example for the model DR-2 + EPOS LHC.
It can be seen that the calculation error introduced by this method is obviously smaller than the measurement error, which shows that the introduction of the SVR is sensible in this analysis.

\begin{figure}[!htbp]
  \centering
  \includegraphics[width=\textwidth]{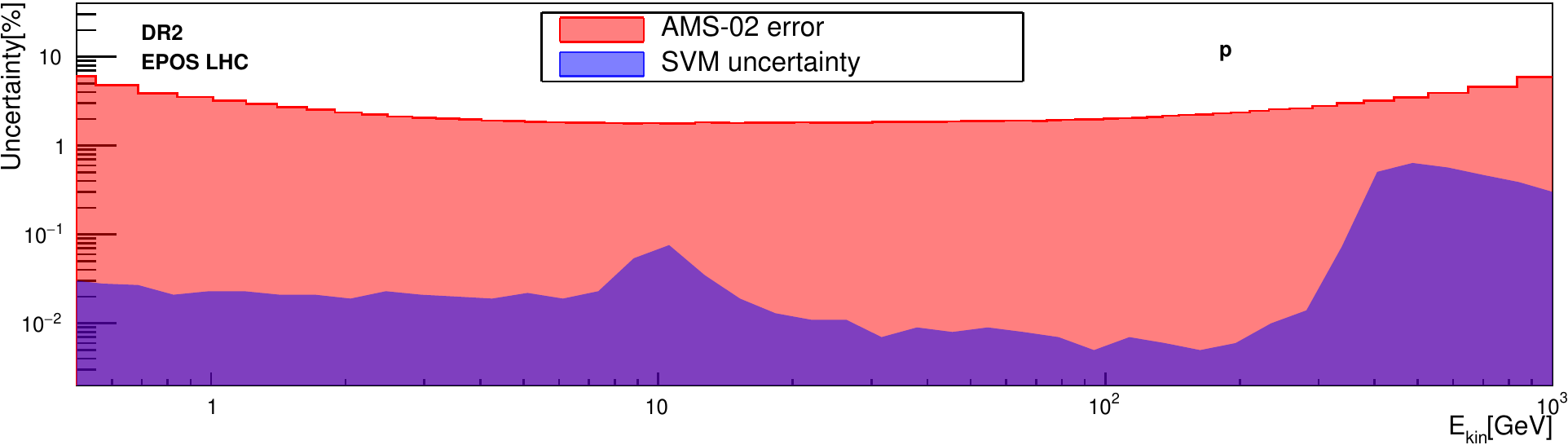}\\
  \includegraphics[width=\textwidth]{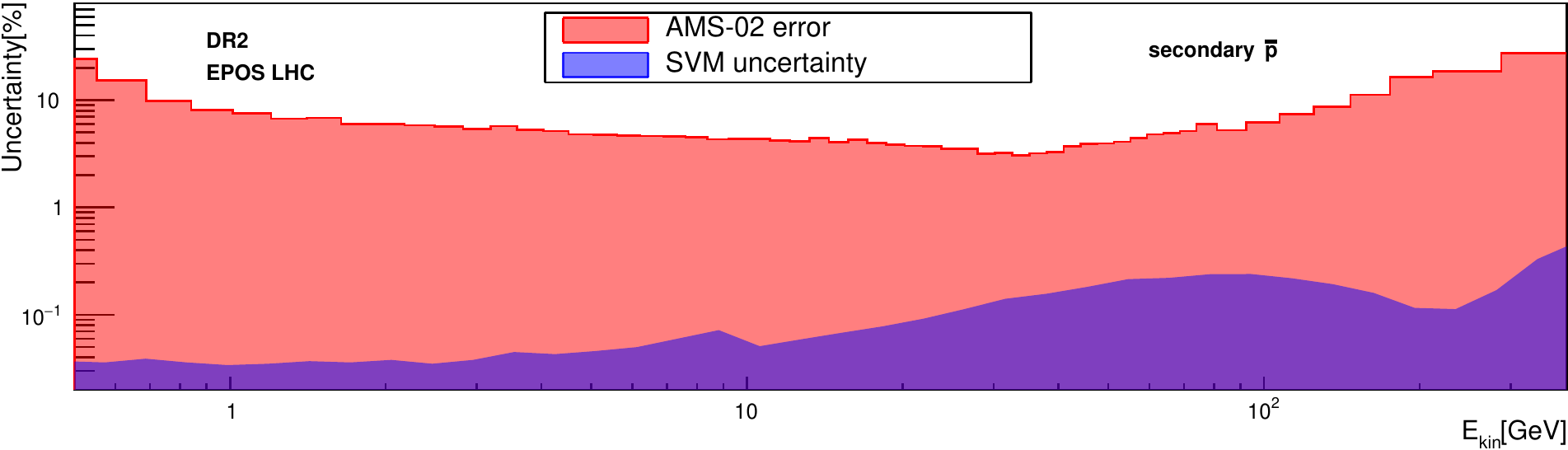}\\
  \includegraphics[width=\textwidth]{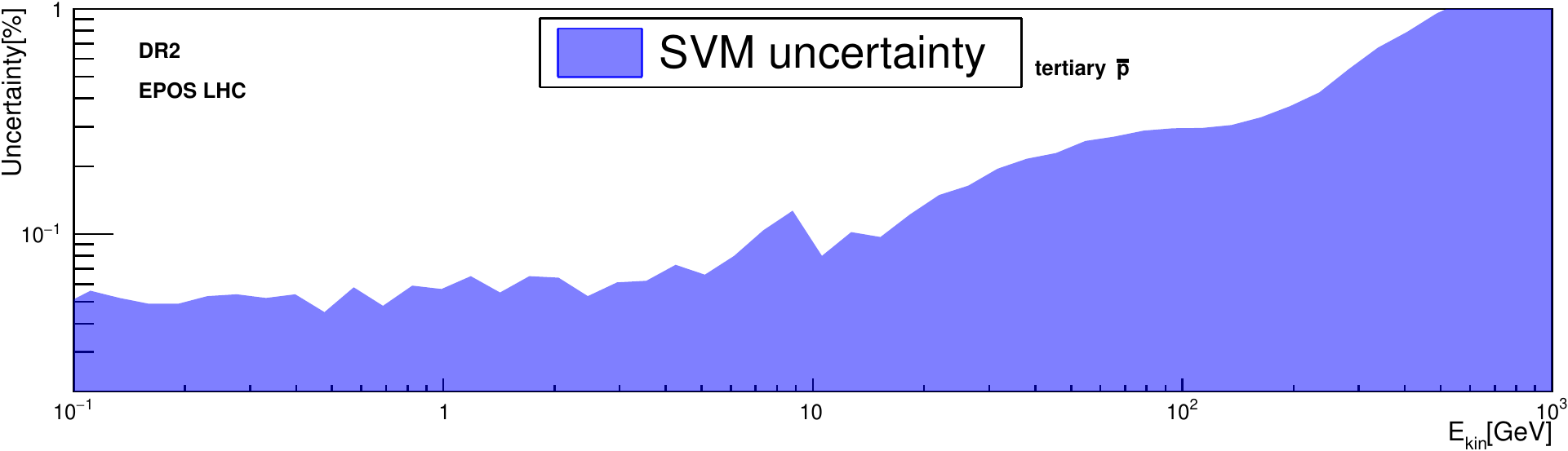}
  \caption{The 1$\sigma$ bands of the relative differences between the predicted CR fluxes and GALPROP results in the DR-2 + EPOS LHC model, compared with the uncertainty of AMS-02 measurements~\cite{Aguilar:2015ooa,Aguilar:2016kjl}. The upper, middle, and lower panels represent for the CR proton, secondary antiproton, and tertiary antiproton, respectively.}
  \label{fig:predict_uncertainty}
\end{figure}

In order to further guarantee the validity, we calculate thousands of random predictions together with the corresponding GALPROP results for the mentioned CR fluxes, and show the 1$\sigma$ uncertainty band of their relative differences in comparison with the AMS-02 measurement uncertainties in Fig.~\ref{fig:predict_uncertainty}.
For both the proton and secondary antiproton fluxes, the differences between the prediction and GALPROP results are always smaller than the measurement uncertainties by one order of magnitude or more.
The tertiary antiprotons that are produced by the interaction between the secondary antiprotons and ISM hold larger relative differences than the other two cases.
Fortunately, these differences would not affect the observed antiproton flux, since the tertiary fluxes are much lower than the secondary fluxes.

Although only the training accuracies for the DR-2 + EPOS LHC model are shown in this subsection, we have ensured that all the models discussed in this work result in similar training accuracies.

\subsection{Training for the DM contribution}

The training procedure for the DM contribution is similar to that in Sec.~\ref{section:training_for_the_background_spectrum} except for two differences.
Firstly, the DM contribution is irrelevant with the injection of CR particles, thus we only need to involve the propagation parameters in the training.
Secondly, the shape of the DM injected spectrum depends on both the annihilation channel and DM mass $m_\chi$, thus we need to separately perform the training for each injection.
In order to approach a more flexible predicting model, we just train a series of fluxes resulting from the injection of delta function in each energy bin, and then linearly combine them according to the DM injected spectrum.
These injection models could only be used to predict the DM contribution, since their spatial distribution is set according to the NFW profile.

\begin{figure}[!htbp]
  \centering
  \includegraphics[width=0.45\textwidth]{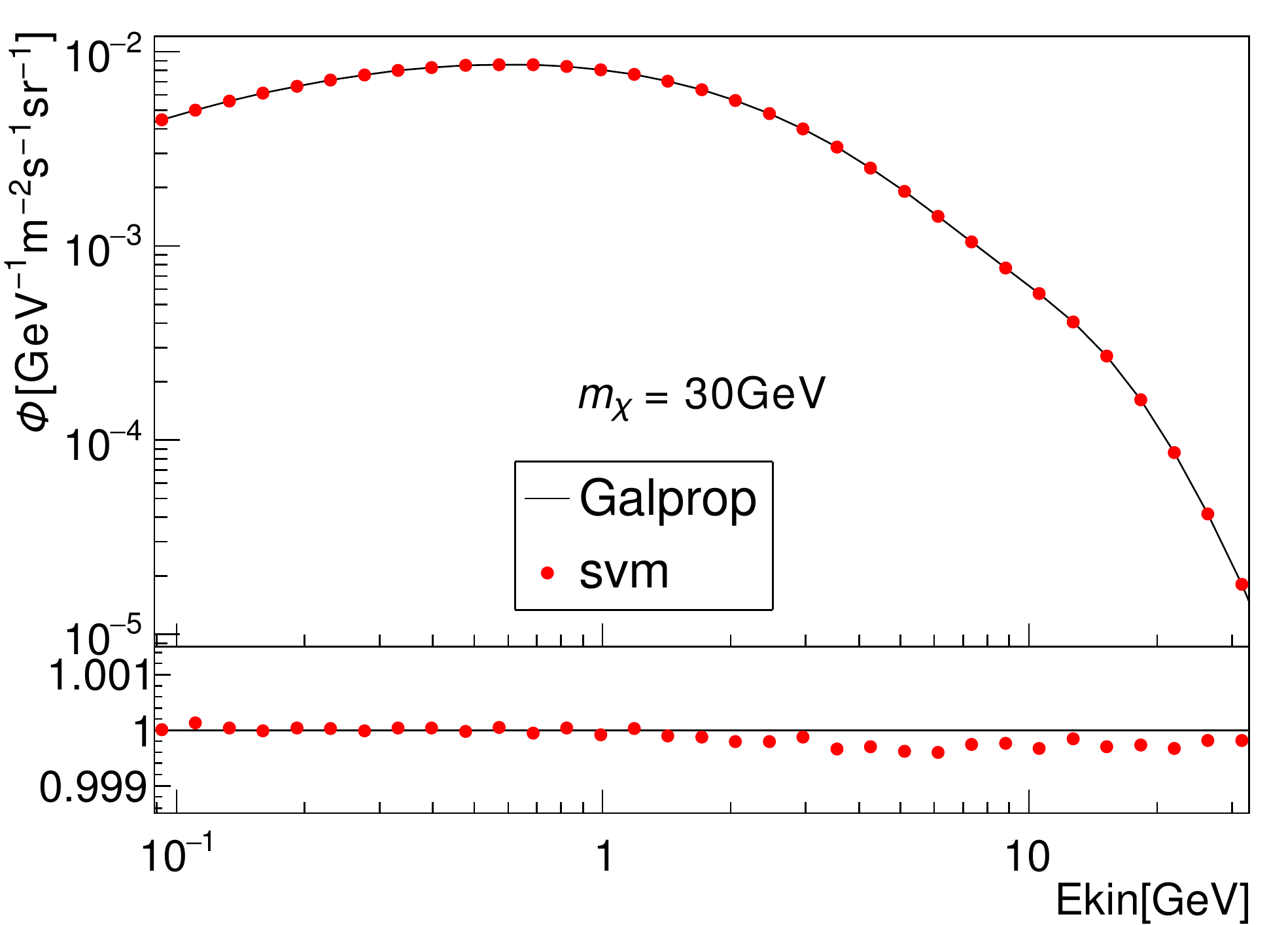}\\
  \caption{An example of the antiproton flux resulting from DM annihilation to $b\bar{b}$ predicted by the SVM models in comparison with the GALPROP result. The DR-2 propagation model along with a randomly specified parameter set is used. The DM mass is taken to be $30\GeV$.}
  \label{fig:DM_antip}
\end{figure}

As a result, We show a random prediction example for the $b\bar{b}$ channel with a annihilating DM mass of $30\GeV$ in the DR-2 propagation model in Fig.~\ref{fig:DM_antip}.
The DM contributions are much smaller than the measurements, thus we do not compare the corresponding relative difference with the measurement uncertainty.
It can be seen that these predictions are even more accurate than the background predictions with relative differences at level of $0.01\%$.
This is because that the predicting models for the DM contribution are simpler as they are constructed with only the propagation parameters.

\section{Implications for DM annihilation} \label{section:implications_for_dm_annihilation}

In this section we determine the confidence regions in the $m_\chi-\langle\sigma v\rangle$ plane favored by the AMS-02 antiproton measurement by scanning the parameter space.
For the 2-dimensional parameter space of $(m_\chi, \langle\sigma v\rangle)$, the boundary of the 1$\sigma$, 2$\sigma$ and 3$\sigma$ regions is given by varying the profile likelihoods as $\chi^2_{1\sigma,2\sigma,3\sigma}=\chi^2_{\mathrm{best}} + 2.29$, 6.16, 11.6, respectively.
We use the package NLOPT~\cite{NLOPTWebSite} to search for the best-fit and the required boundaries in the $m_\chi-\langle\sigma v\rangle$ plane.

In order to find the corresponding profile likelihood for each given point $(m_\chi, \langle\sigma v\rangle)$, we optimize throughout the 2$\sigma$ confidence region of the astrophysical parameters constrained by the $\mathrm{B}/\mathrm{C}$, $\mathrm{Be}^{10}/\mathrm{Be}^9$ and proton data.
Using the predicting models built in the above section, we could determine whether a set of propagation + injection + modulation parameters $\bm{\theta}$ falls inside the 2$\sigma$ confidence region, and reproduce the antiproton flux accurately within a time scale of $\mathcal{O}(0.1)$s.
This optimization limited in the constrained astrophysical parameter region is also approached using the NLOPT with the augmented Lagrangian algorithm~\cite{Birgin:2008iuc}.
The modulation potential for negative charged particles $\phi$ is allowed to vary from 0.5 to 1.5 times of $\phi_+$ as mentioned in Sec.~\ref{section:solar_modulation}.
Since the uncertainties associated with the heavy nuclear reaction cross sections, such as $\mathrm{C, O \rightarrow B}$, would affect the $\mathrm{B}/\mathrm{C}$ prediction and the determination of the propagation parameters, we also introduce a free scale factor $c_{\bar{p}}$
for the antiproton flux following Ref.~\cite{Lin:2016ezz}.

\begin{figure}[!htbp]
  \centering
  \includegraphics[width=0.45\textwidth]{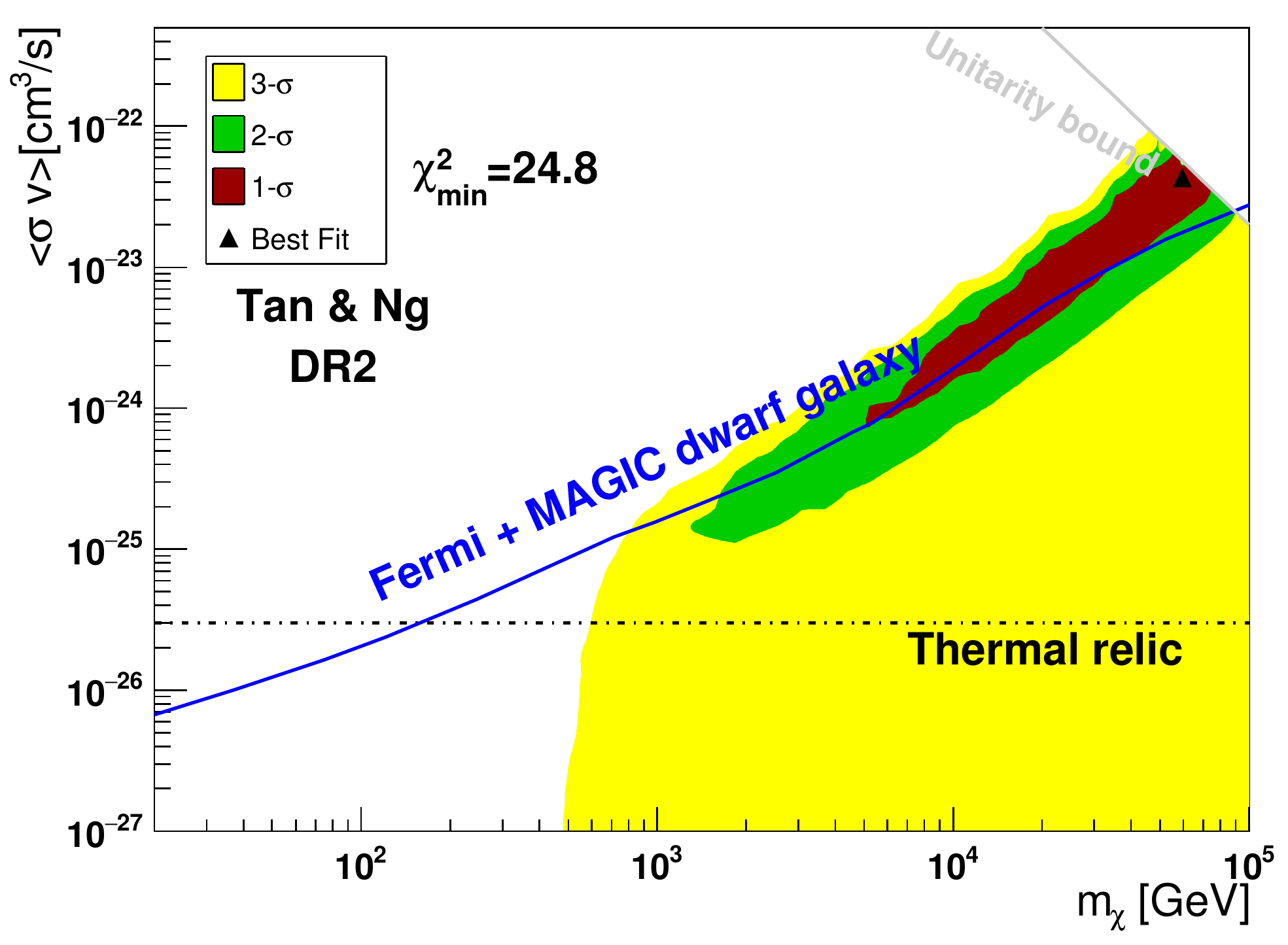}
  \includegraphics[width=0.45\textwidth]{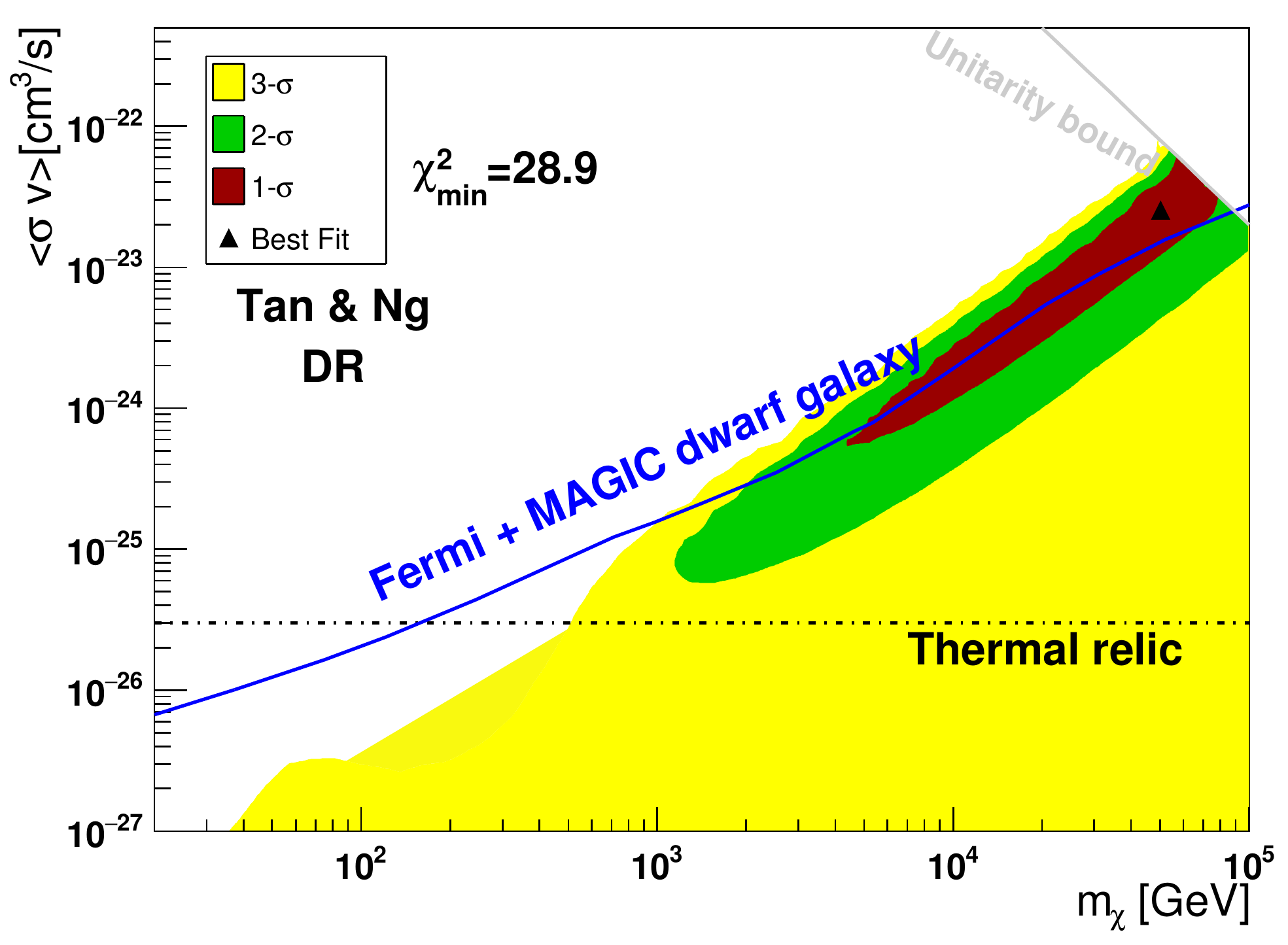}\\
  \includegraphics[width=0.45\textwidth]{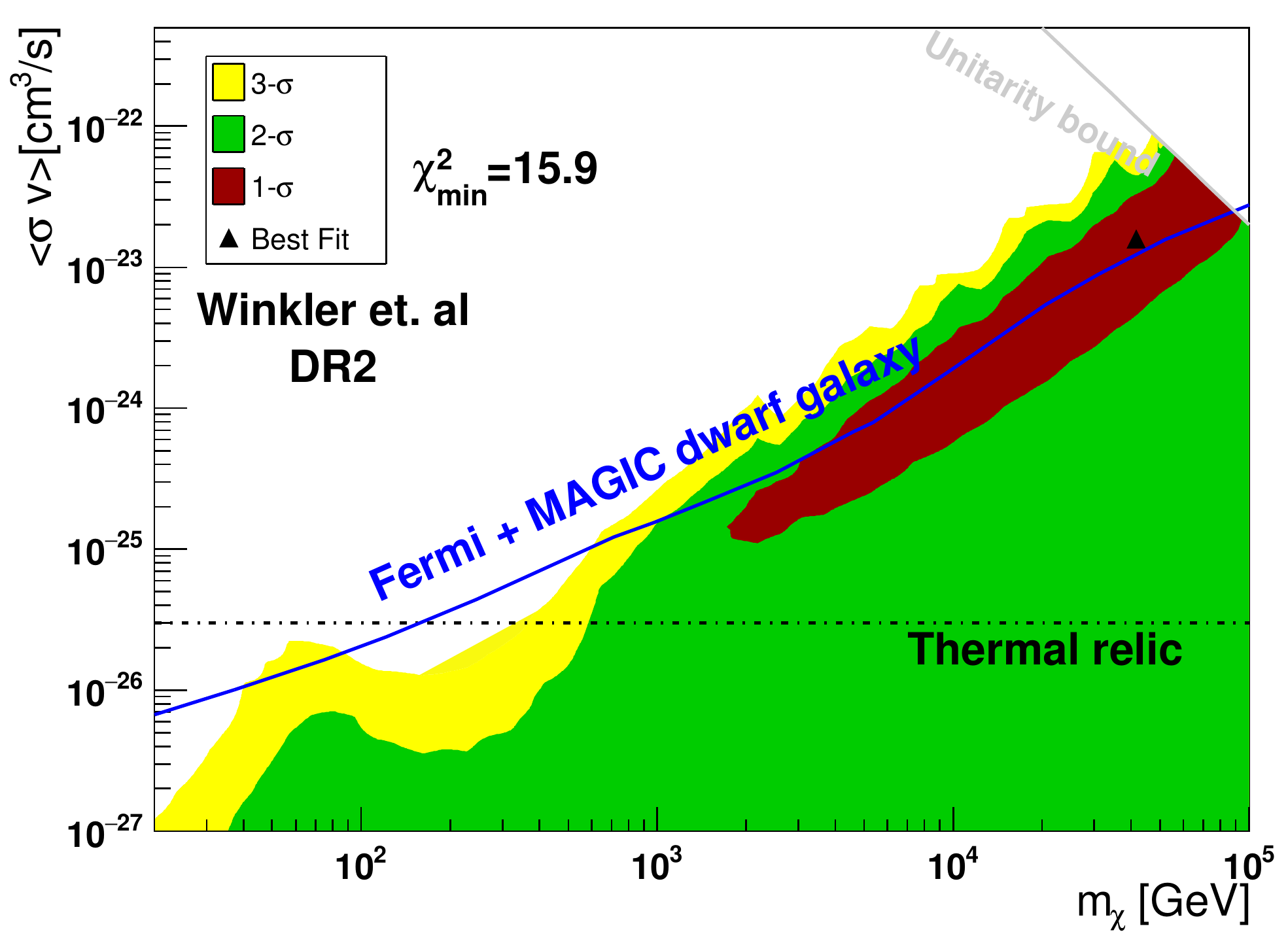}
  \includegraphics[width=0.45\textwidth]{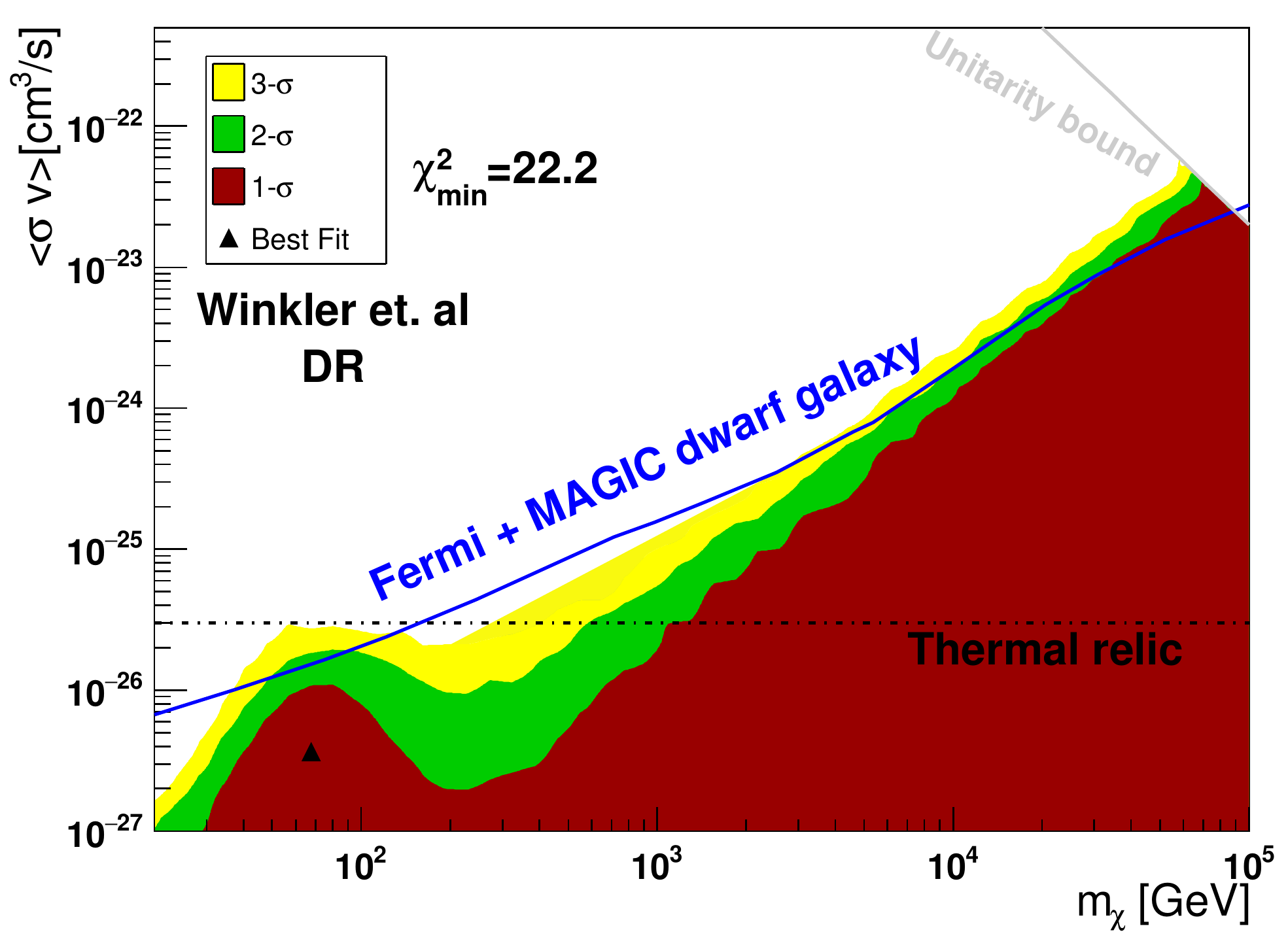}\\
  \includegraphics[width=0.45\textwidth]{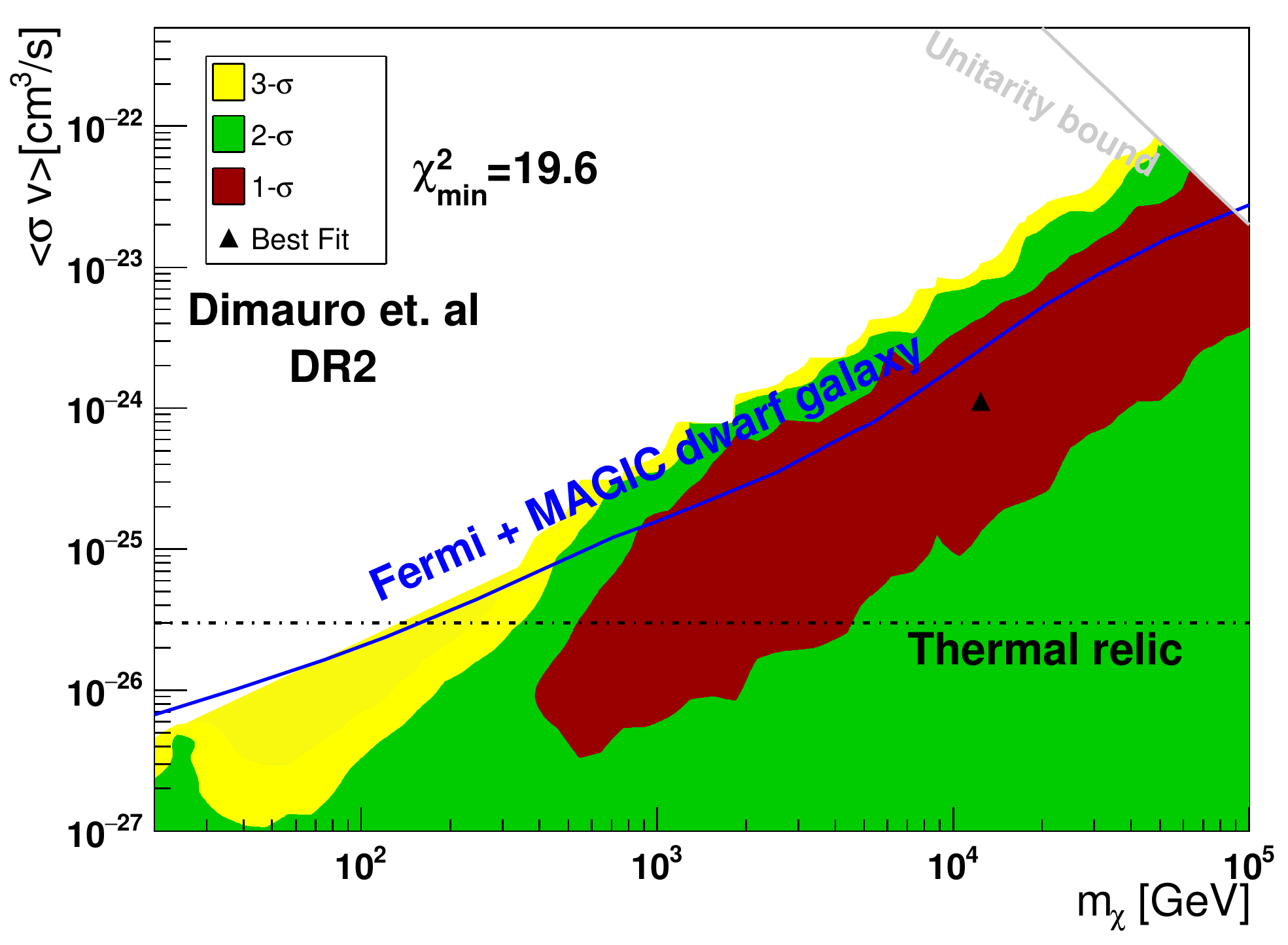}
  \includegraphics[width=0.45\textwidth]{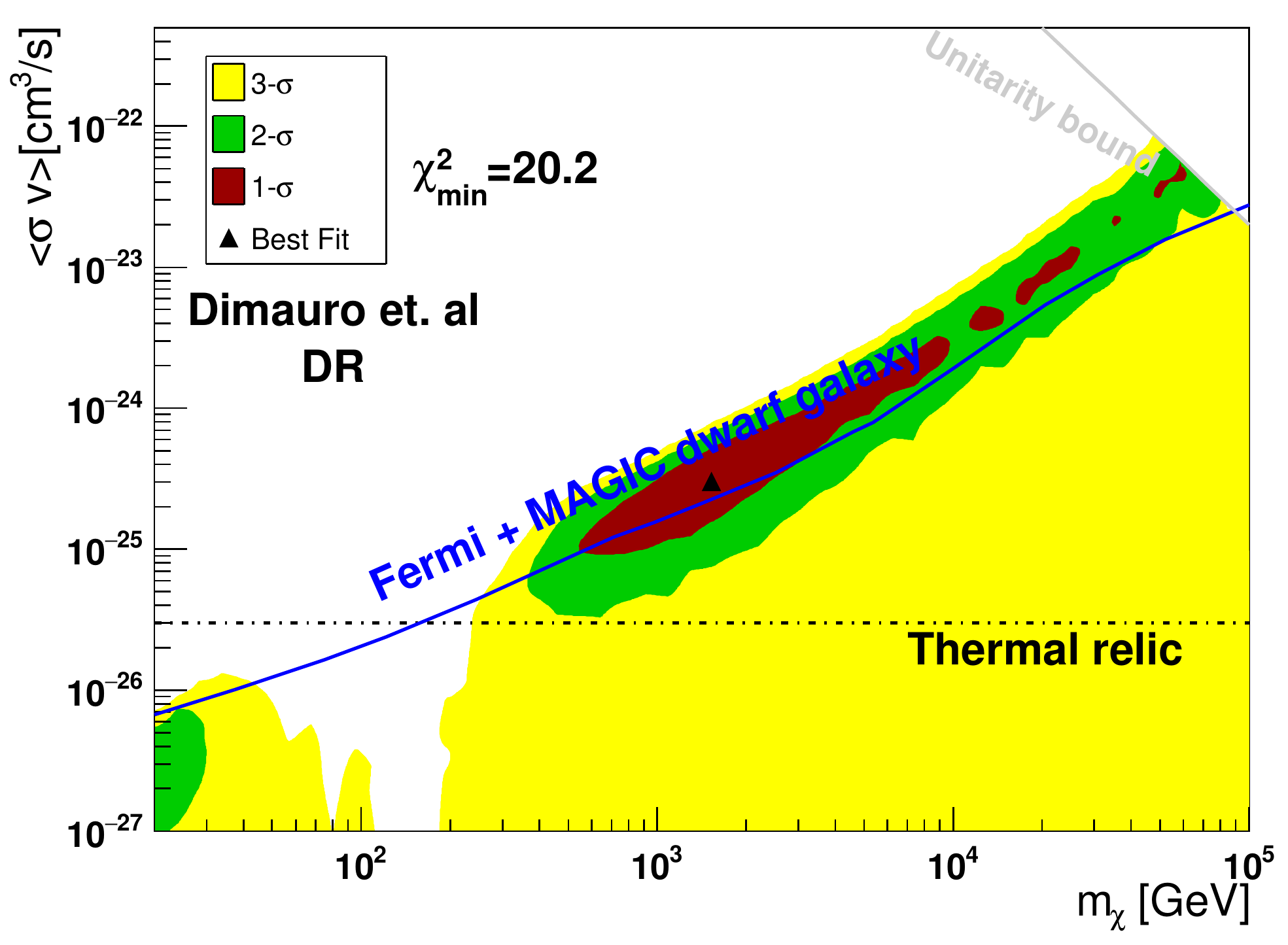}\\
  \caption{The 1$\sigma$, 2$\sigma$ and 3$\sigma$ confidence regions in the $m_\chi-\langle\sigma v\rangle$ plane favored by fitting  to the AMS-02 antiproton flux~\cite{Aguilar:2016kjl}, for the hadronic models from Tan \& Ng~\cite{1983JPhG....9.1289T} (upper panels), Winkler~\cite{Winkler:2017xor} (middle panels) and di Mauro et. al~\cite{diMauro:2014zea} (lower panels).
    The left and right panels represent the results for the DR-2 and DR propagation models, respectively.
    The unitarity bound~\cite{Griest:1989wd} and the limit of dwarf galaxies $\gamma$-ray emission from Fermi + MAGIC~\cite{Ahnen:2016qkx} is also plotted.
}
  \label{fig:analytical_result}
\end{figure}

\begin{figure}[!htbp]
  \centering
  \includegraphics[width=0.45\textwidth]{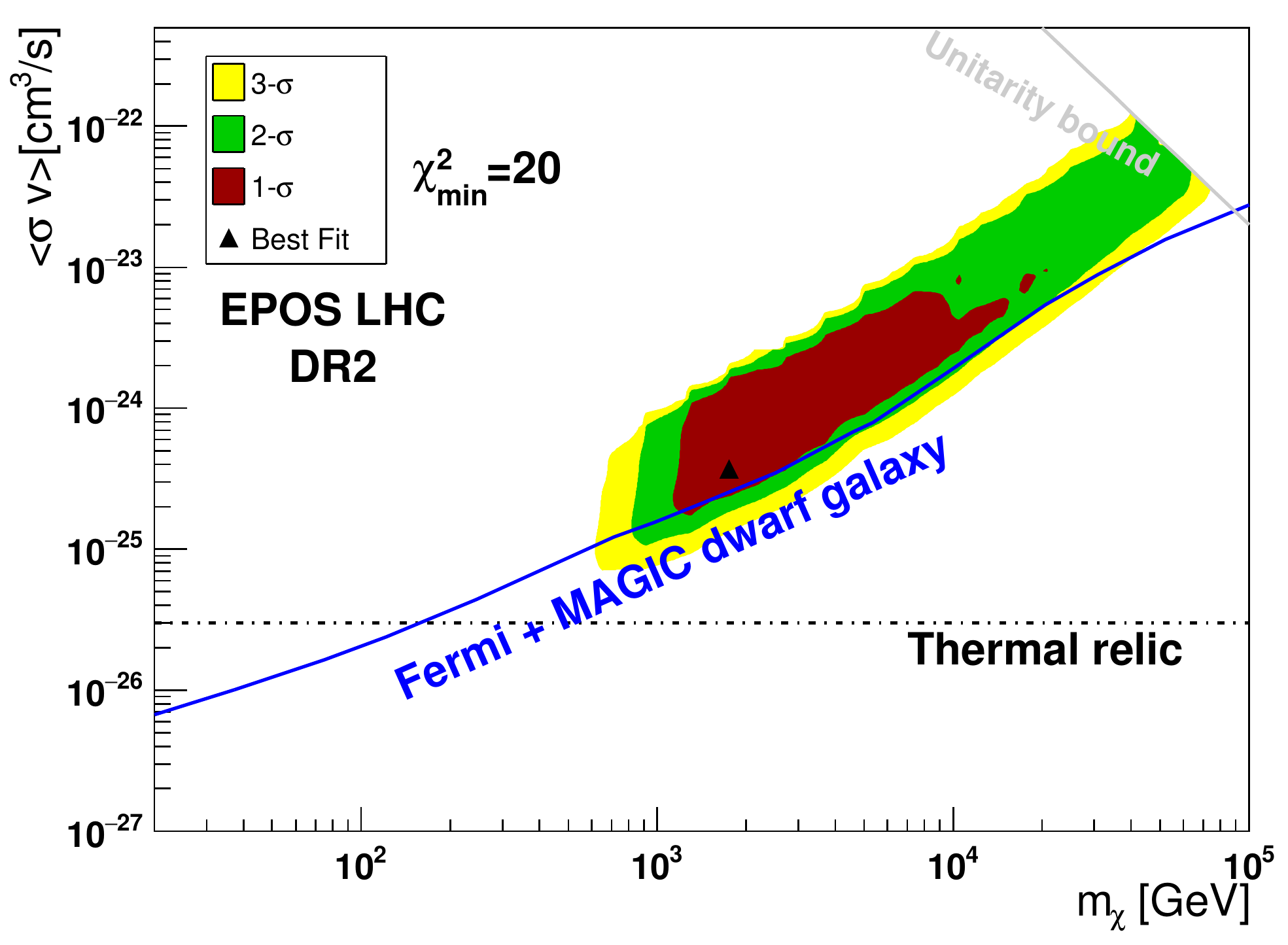}
  \includegraphics[width=0.45\textwidth]{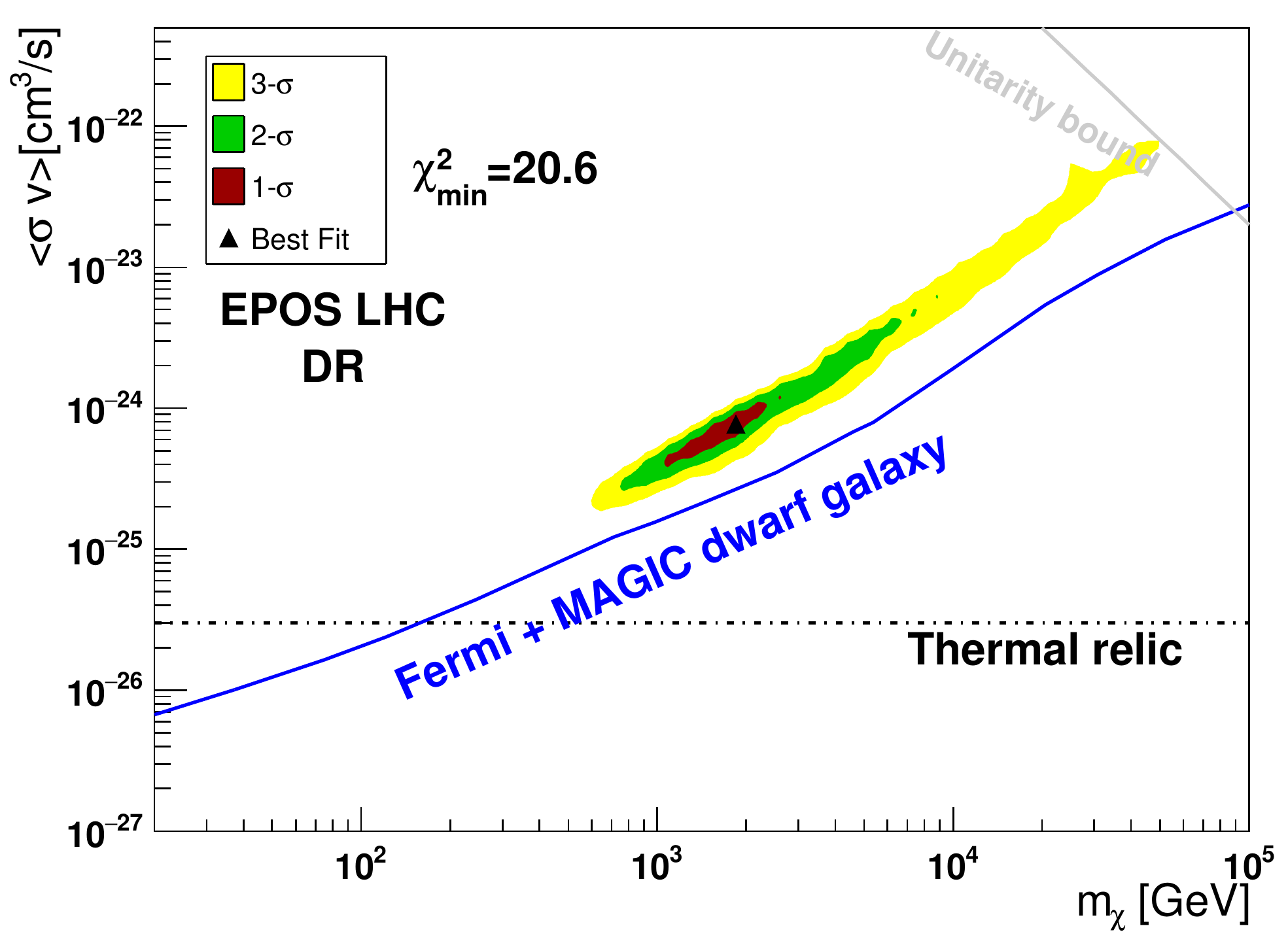}\\
  \includegraphics[width=0.45\textwidth]{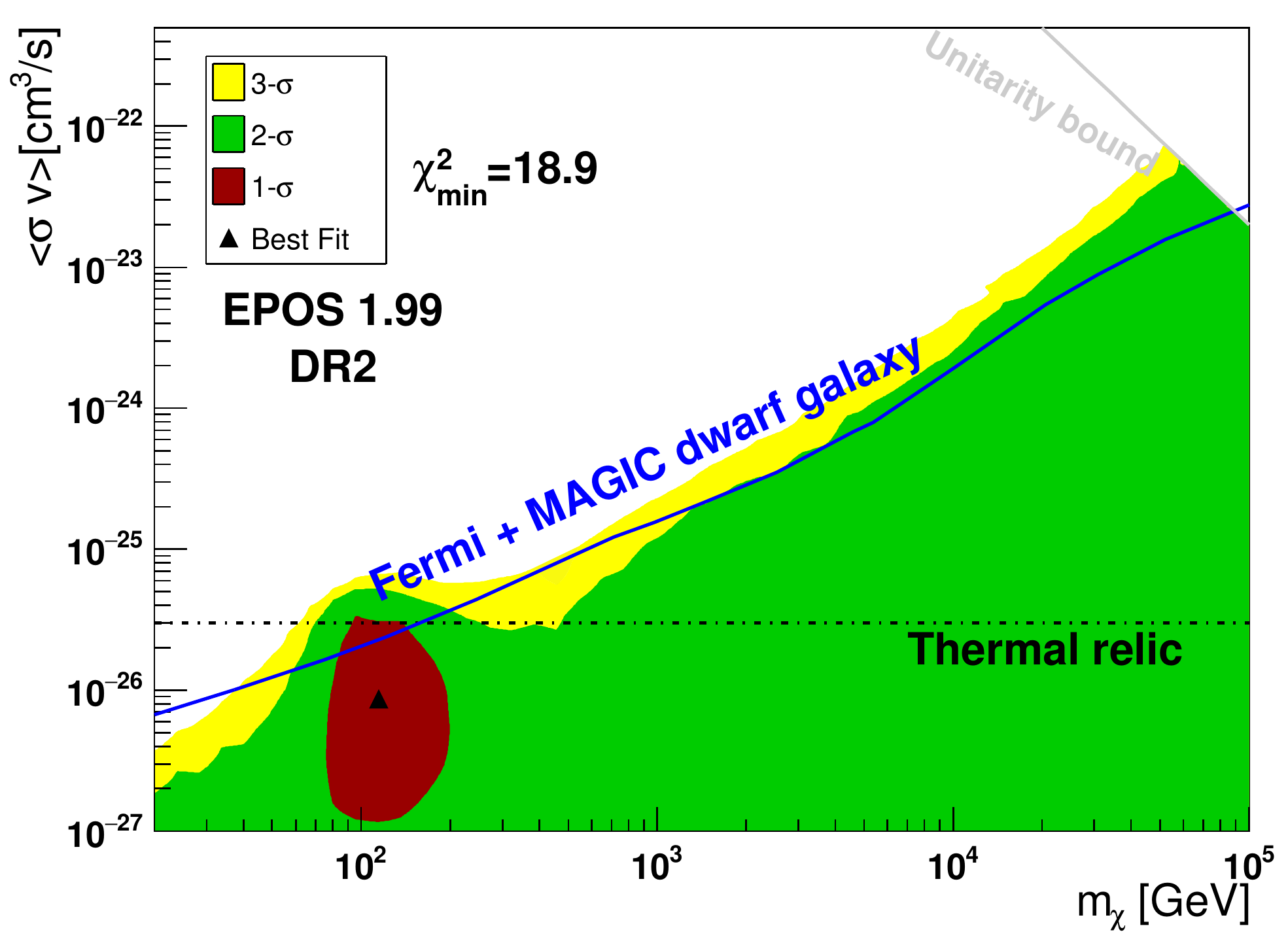}
  \includegraphics[width=0.45\textwidth]{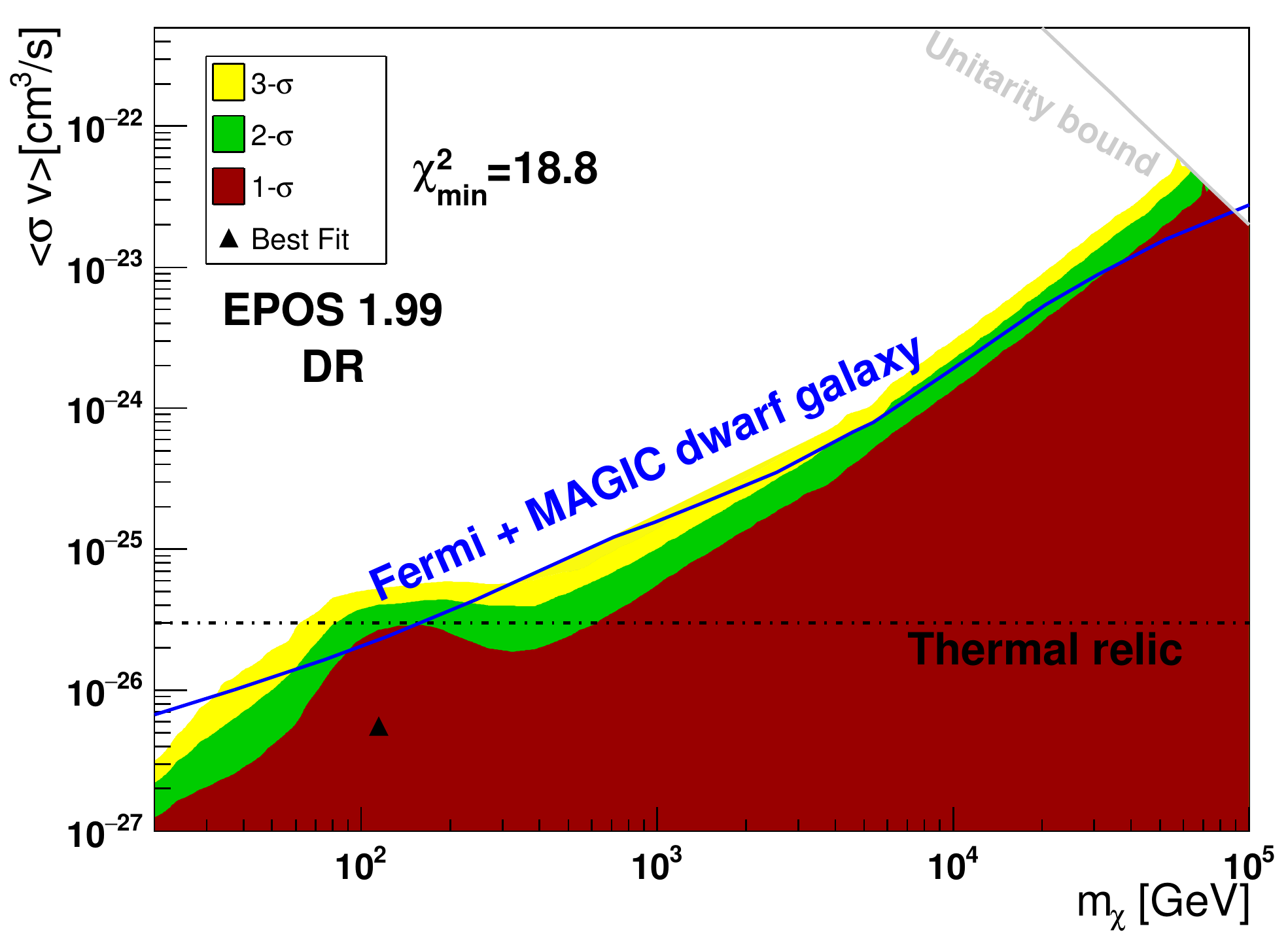}\\
  \caption{The same as Fig.~\ref{fig:analytical_result}, but for the EPOS LHC and EPOS 1.99 hadronic models.}
  \label{fig:generator_result}
\end{figure}

We investigate the DM implication in all the chosen hadronic models and propagation models.
The results for the empirical and phenomenological hadronic models are shown in Fig.~\ref{fig:analytical_result} and Fig.~\ref{fig:generator_result}, respectively.
We consider both the DR and DR-2 propagation models in these figures.
The different colors in the figures represent the $1\sigma, 2\sigma, 3\sigma$ regions favored by the AMS-02 data.
These regions that are not closed but extend to zero indicate upper limits at the corresponding significance instead of the interpretation of DM signatures.

We find that the differences arising from hadronic models are always larger than those from propagation models.
It can be clearly seen from these figures that in most cases the significance of a DM signal is smaller than $2\sigma$.
Among all the hadronic models, only the EPOS LHC model requires a DM contribution with $m_\chi\sim1\TeV$ and $\langle\sigma v\rangle\sim5\times10^{25}\cm^3/\sec$ at a confidence level over $99.7\%$, and result in an enclosed 3$\sigma$ confidence region.
Unfortunately the whole 3$\sigma$ confidence region is excluded by the combined limit of the dwarf galaxy $\gamma$-ray emission set by Fermi and Magic~\cite{Ahnen:2016qkx}.
However, the upper limit on the DM annihilation rate by the $\gamma$-ray emission from dwarf galaxies is model dependent, such as the case discussed in Ref.~\cite{Xiang:2017jou}.
If the DM annihilation is velocity dependent, it is still possible to contribute to the AMS-02 antiproton flux.
The significance of the DM contribution for the other models is less than 2$\sigma$.
For the Winkler's model and the EPOS 1.99 model, the confidence regions expand throughout the lower panel and their upper boundaries become upper-limits on the DM annihilation rate.

A bump around $100\GeV$ is found in the upper boundaries for the EPOS 1.99 and Winkler's model due to the underestimation of the secondary CR antiproton below $\sim10\GeV$.
This result is different from Refs.~\cite{Cui:2018nlm,Cuoco:2019kuu} in which the underestimation is quite significant and an enclosed contour instead of a bump is found.
The reason is that those two works take the DR propagation model and adopt a charge independent solar modulation potential.

\begin{figure}[!htbp]
  \centering
  \includegraphics[width=0.45\textwidth]{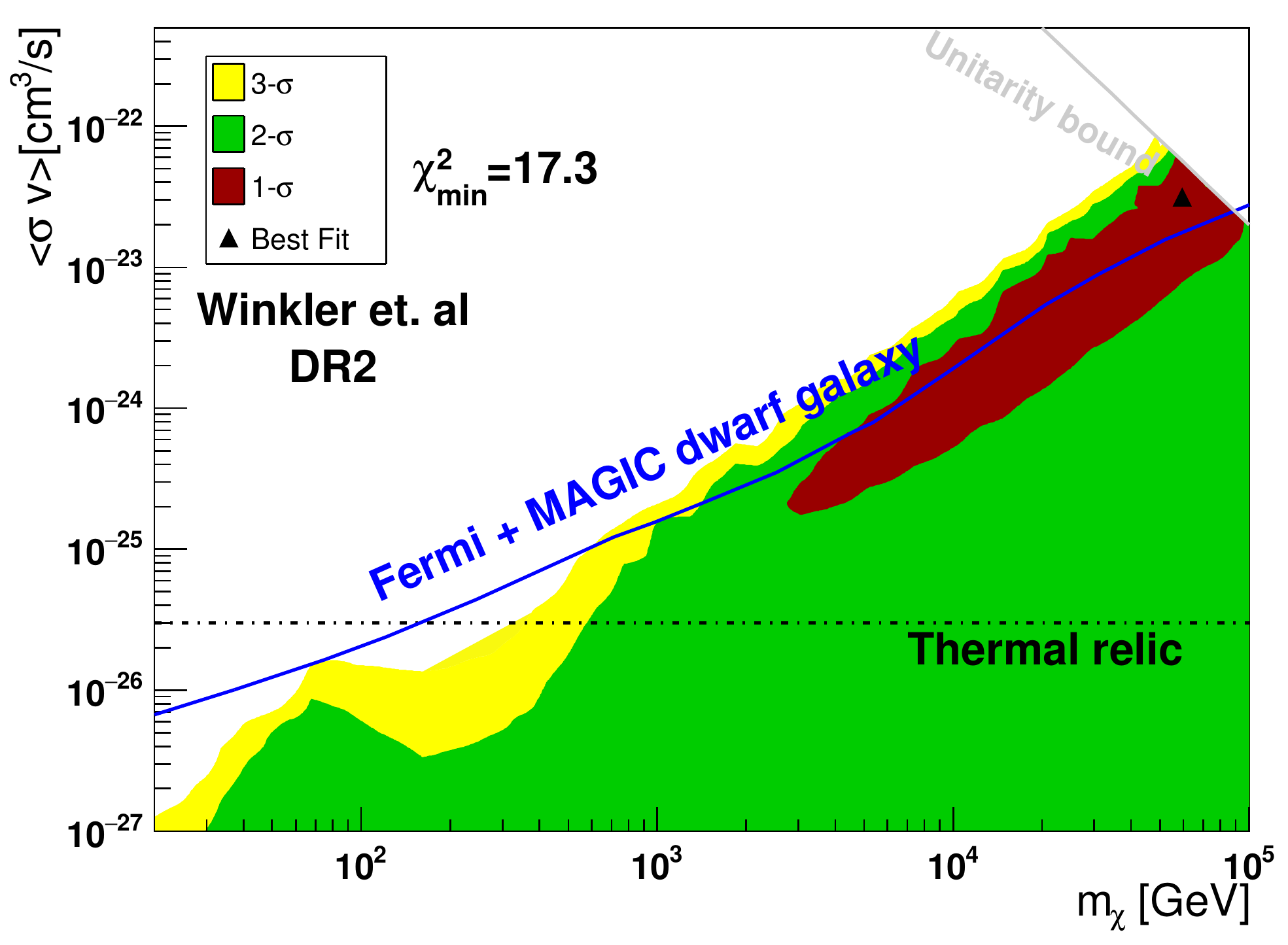}
  \includegraphics[width=0.45\textwidth]{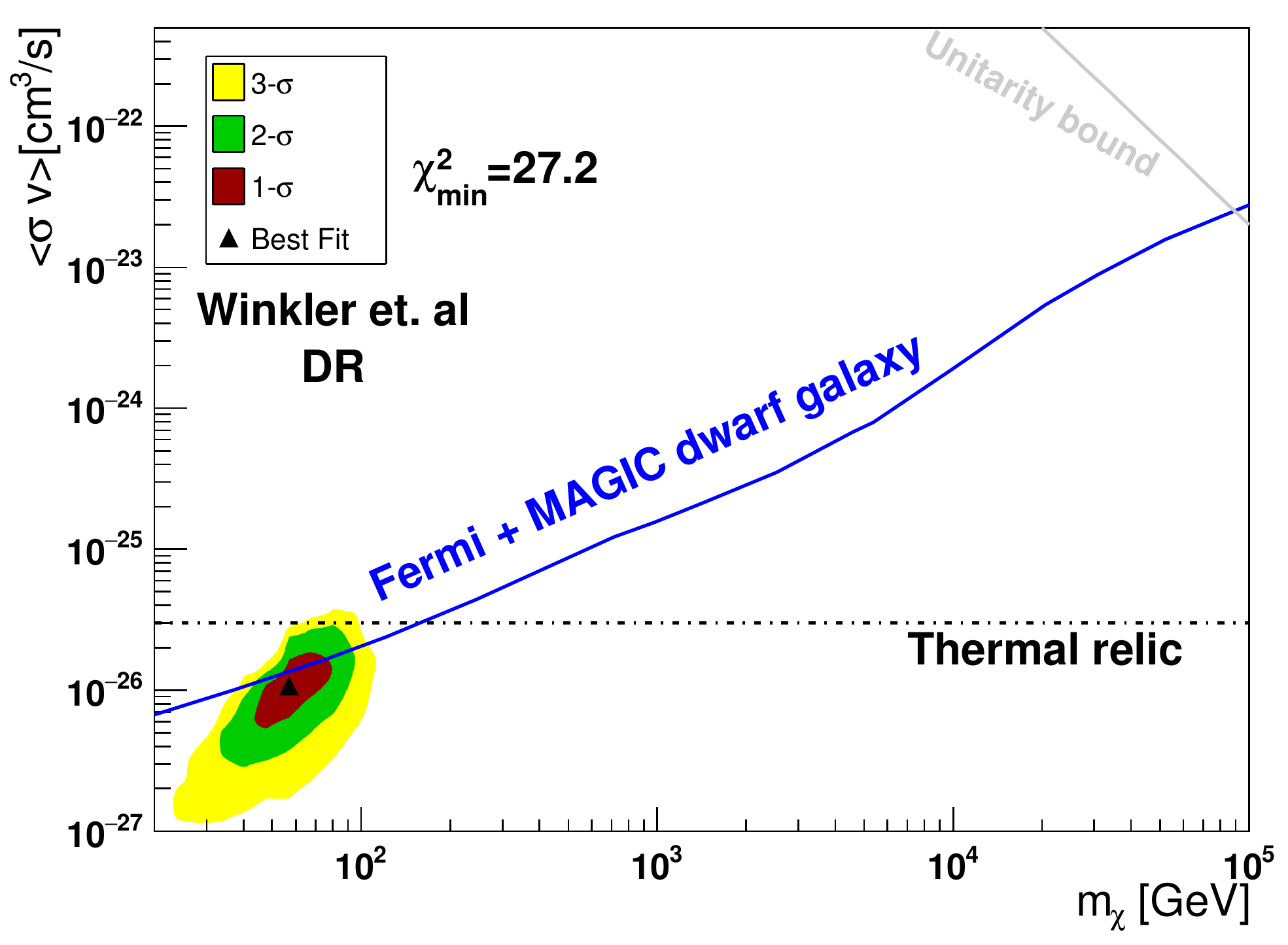}\\
  \caption{Similar to Fig.~\ref{fig:analytical_result}, but a charge independent FFA modulation potential is used in the calculation for the hadronic model from Winkler.}
  \label{fig:modulation_variation}
\end{figure}

For comparison, we also show the confidence regions derived under the assumption $\phi_-=\phi_+$ for the Winkler's model in Fig.~\ref{fig:modulation_variation}.
It can be seen that the confidence region for the DR-2 model stays unchanged while that for the DR model becomes an enclosed contour around $100\GeV$.
The DR model tends to underestimate the secondary CR antiprotons at low energies, which is a well known fact and has been discussed very early~\cite{Moskalenko:2002yx,Trotta:2010mx,Hooper:2014ysa}.
The deficiency of antiprotons at low energy is one of the main reasons to propose the DR-2 model~\cite{Evoli:2011id}.
Our study shows that this deficiency can be corrected by a charge dependent solar modulation.
A question is whether our allowing range for the negative charged potential ($0.5\phi_+\sim1.5\phi_+$) is suitable for the charge-sign dependent modulation effect.
Different simplifications on the solar modulation do affect the implication of the DR models.
For instance, adopting a parametrized modulation potential to describe the charge-sign dependent effect, Cholis et. al. have found that the deficiency is still significant~\cite{Cholis:2019ejx}.
A thorough solution to this problem is numerically solving the transport equation in the Solar system~\cite{Maccione:2012cu,Potgieter:2014pka,Kappl:2015hxv} instead of the FFA method.
Such study is out of scope of this paper and is left for the future work.

In all, the significance of the DM contribution in the antiproton flux depends on the hadronic models. Only the EPOS LHC model would significantly expect the DM contribution at high energies.
Note that the hadronic models are being developed with the collection of new collider data.
For instance, a measurement of the antiproton production in $p-\mathrm{He}$ collisions at $\sqrt{s}=110\GeV$ has been recently presented by the LHCb collaboration~\cite{Aaij:2018svt}, which shows a deviation from the predictions of many current hadronic models.
Korsmeier et al. have developed the model of di Mauro et al. to fit this measurement~\cite{Korsmeier:2018gcy}.
In the future, more relevant collider data will be collected and the hadronic models discussed above would be updated.
The implication for the DM contribution in the CR antiproton flux will become more clear.

\section{Conclusions} \label{section:conclusions}

In this work, we used the SVR method to rapidly predicted the CR antiproton fluxes with either the astrophysical or DM origination, and adopted them to investigate the DM parameter regions favored by the AMS-02 antiproton measurement.
In the implementation, we have ensured that these predicted antiproton fluxes are accurate approximations of the GALPROP solutions in comparison with the current measurement uncertainties.
The uncertainties of the astrophysical parameters have been well taken into account in the scanning analysis.

The DM implications in several propagation and hadronic models have been separately investigated.
These models would occasionally underestimate the antiproton fluxes below several $\GeV$ or around $100\GeV$, and leave the room for the DM contribution with a DM mass around $100\GeV$ or $1\TeV$, respectively.
If the charge-sign dependent solar modulation is taken into account, the favored region for a DM mass around $\sim100\GeV$ could become not significant in comparison with the results in Refs.~\cite{Cui:2018nlm,Cuoco:2019kuu}.
However, this conclusion depends on the simplified charge-sign dependent solar modulation model used in our analysis.
Further study based on the numerical solar modulation solution is needed to verify it.
On the other hand, the implication for the $\TeV$ DM depends on the hadronic models.
The DM contribution with a mass of $\TeV$ is only favored in the EPOS LHC model with a significance over 3$\sigma$.
The future collider measurement will help to update these hadronic models and leave us a more certain conclusion on the DM implication in the CR antiprotons.

\begin{acknowledgments}
This work is supported by the National Key R\&D Program of China (No.~2016YFA0400200),
the National Natural Science Foundation of China (Nos.~U1738209 and 11851303).
\end{acknowledgments}

\appendix
\section{Support-vector machine} \label{section:SVM}

The SVM algorithm was originally invented by V. N. Vapnik and A. Y. Chervonenkis in 1963 as a binary linear classifier.
Given a dataset of $n$ points in $l$ dimensional space $\{ (\mathbf{x}_i, y_i) | i = 1, 2 \dots n \}$ where the $y_i$s are either 1 or -1, the original SVM algorithm would tend to find two parallel boundary hyperplanes that could separate the two kinds of points with a maximum margin.
These points are not promised to be linearly separable, thus a ``soft-margin'' rather than the real margin is adopted in the searching.
The algorithm could then mathematically expressed as
\begin{equation}
  \inf\left\{\frac{1}{n}\sum_{i=1}^n\max(0, 1 - y_i(\bm{\omega}\cdot\bm{x}_i - b) + \lambda|\bm{\omega}|^2
    \middle| \bm{\omega} \in \mathcal{R}^l, b \in \mathcal{R}  \right\},
  \label{eq:SVM_soft_margin}
\end{equation}
where $\lambda$ is a positive number determined manually.
When the points are linearly separable, it is able to find a suitable $\bm{\omega}$ that keep all the $\max(0, 1 - y_i(\bm{\omega}\cdot\bm{x}_i - b))$ equal to 0, and $\bm{\omega}$ at the infimum in Eq.~\ref{eq:SVM_soft_margin} naturally corresponds to the maximum margin.
When they are not separable, the sum would always be positive, and $\lambda$ is used to trade-off between the classification error and the margin size.
At the final solution, only a part of points that satisfy $1 - y_i(\bm{\omega}\cdot\bm{x}_i - b) > 0$ would directly contribute to the value of $\bm{\omega}$ and $b$.
These points are called as "support-vector".

Based on the SVM algorithm, the SVR algorithm was then introduced in 1996~\cite{Drucker:1996svr}
\begin{equation}
  \inf\left\{U\sum_{i=1}^n\max\left(\varepsilon, |\bm{\omega}\cdot\bm{x}_i + b - y_i|\right) + \frac{1}{2}|\bm{\omega}|^2
  \middle| \bm{\omega} \in \mathcal{R}^l, b \in \mathcal{R}  \right\},
  \label{eq:SVR}
\end{equation}
where the value of $y_i$ is no longer dispersed, and the positive $\varepsilon$ is introduced to control the tolerance between $y_i$ and the predicted value $\bm{\omega}\cdot\bm{x}_i + b$.
Smaller $\varepsilon$ would lead to more support-vectors and vise versa.

Eq.~\ref{eq:SVR} is a linear regression; in order to deal with the non-linear problems, a kernel trick is introduced~\cite{Boser92atraining}.
Any non-linear function $\hat{y}(\bm{x})$ could be transformed to a linear one by projecting the input parameter $\bm{x}$ to a space with  higher dimension using a transform $\hat{\bm{\varphi}}(\bm{x})$, such that
\begin{equation}
  \hat{y}(\varphi) = \bm{W}\cdot\bm{\varphi} + b\phantom{jjjjj}(\bm{\varphi}\in \mathcal{R}^k,\phantom{j}k > l).
  \label{eq:high_dimension}
\end{equation}
The kernel function is defined by the transform $\hat{\bm{\varphi}}(\bm{x})$ as
\begin{equation}
  k(\bm{x}_i, \bm{x}_j) \equiv \hat{\bm{\varphi}}(\bm{x}_i) \cdot \hat{\bm{\varphi}}(\bm{x}_j).
  \label{eq:kernel_function}
\end{equation}
In the implementation, we could directly adopt the kernel function to avoid the complicated calculation of $\hat{\bm{\varphi}}(\bm{x})$ when calculating the dot product in the transformed space $\mathcal{R}^k$.
In this work, we adopt the widely used Gaussian radial basis function (RBF)
\begin{equation}
  k(\bm{x}_i, \bm{x}_j) = \exp\left(-\gamma|\bm{x}_i - \bm{x}_j|^2\right),
  \label{eq:rbf_kernel}
\end{equation}
where $\gamma$ is a positive free parameter.
The RBF corresponds to a transformed space with infinite dimension~\cite{Shashua:2009itm}, thus it could be used to deal with almost all kinds of target functions.

Therefore, the SVR for non-linear problem is
\begin{equation}
  \inf\left\{U\sum_{i=1}^n\max\left(\varepsilon, | k(\bm{\omega},\bm{x}_i) + b - y_i|\right) + \frac{1}{2} k(\bm{\omega}, \bm{\omega})
  \middle| \bm{\omega} \in \mathcal{R}^l, b \in \mathcal{R}  \right\}.
  \label{eq:SVR_nonlinear}
\end{equation}
When the dataset size $n$ is large, directly searching for the infimum in Eq.~\ref{eq:SVR_nonlinear} would require us to calculate the kernel function for many times.
In order to avoid the time-consuming calculation for the kernel function, the problem is always transformed to its Lagrangian dual before being solved.
We can express Eq.~\ref{eq:SVR_nonlinear} as
\begin{equation}
  \begin{aligned}
    \text{minimize} & & U\sum_{i=1}^N\xi^*_i + U\sum_{i=1}^N\xi_i + \frac{1}{2}k(\bm{\omega}, \bm{\omega}) \\
    \text{subject to} & & \left\{\begin{aligned}
        y_i - k(\bm{\omega}, \bm{x}_i) - b \le& \varepsilon + \xi^*_i \\
        -y_i + k(\bm{\omega}, \bm{x}_i) + b \le& \varepsilon + \xi_i \\
        \xi^*_i \ge 0\phantom{jjjj}\xi_i \ge 0 &\\
      \end{aligned}
      \right.
    \end{aligned}.
  \label{eq:SVR_nonlinear_xi}
\end{equation}
The corresponding Lagrangian is then
\begin{equation}
  \begin{aligned}
    \mathcal{L}=&\frac{1}{2}\bm{W} \cdot \bm{W} + U\sum_{i=1}^N(\xi_i^* + \xi_i)
     - \sum_{i=1}^N\alpha_i^*(-y_i + \bm{W}\cdot \hat{\bm{\varphi}}(\bm{x}_i) + b + \varepsilon + \xi_i^*) \\
    &- \sum_{i=1}^N\alpha_i(y_i - \bm{W}\cdot\hat{\bm{\varphi}}(\bm{x}_i) - b + \varepsilon + \xi_i)
    - \sum_{i=1}^N\gamma_i(\xi_i^* + \xi_i),
  \end{aligned}
  \label{eq:Lagrangian}
\end{equation}
where $\bm{W}=\hat{\bm{\varphi}}(\bm{\omega})$.
The minimization that we search for is just the saddle point of $\mathcal{L}(\bm{W}, b, \xi^*, \xi, \alpha^*, \alpha, \gamma)$ with respect to $\bm{W}$, $b$, $\xi^*$ and $\xi$, in which $\partial\mathcal{L}/\partial\bm{W} = \partial\mathcal{L}/\partial b = \partial\mathcal{L}/\partial\xi=\partial\mathcal{L}/\partial\xi^*=0$~\cite{Shashua:2009itm}.
Substituting these expressions back into Eq.~\ref{eq:Lagrangian}, we could finally reach the dual problem
\begin{equation}
  \begin{aligned}
    \text{minimize     }&  \varepsilon\sum_{i=1}^n(\alpha^*_i + \alpha_i) + \sum_{i=1}^ny_i(\alpha_i-\alpha^*_i) - \frac{1}{2} \sum_{i,j=1}^N(\alpha_i - \alpha^*_i)k(\bm{x}_i, \bm{x}_j)(\alpha_j - \alpha^*_j) \\
    \text{subject to     }& \left\{\begin{aligned}
        &\alpha_i = \alpha_i^*\\
        &0 \le \alpha_i,\alpha_i^* \le U
    \end{aligned}\right.
  \end{aligned}.
  \label{eq:dual_SVR}
\end{equation}
The approximate function for $\hat{y}(\bm{x})$ is then $\hat{h}(\bm{x}) = \sum_{i=1}^n(\alpha^*_i - \alpha_i)k(\bm{x}_i, \bm{x}) + b$.
This dual problem is a simple quadratic programming problem that could be solved fast with the sequential minimal optimization (SMO)~\cite{Platt98sequentialminimal}.

\bibliography{antiproton_svm}

\begin{thebibliography}{92}%
\makeatletter
\providecommand \@ifxundefined [1]{%
 \@ifx{#1\undefined}
}%
\providecommand \@ifnum [1]{%
 \ifnum #1\expandafter \@firstoftwo
 \else \expandafter \@secondoftwo
 \fi
}%
\providecommand \@ifx [1]{%
 \ifx #1\expandafter \@firstoftwo
 \else \expandafter \@secondoftwo
 \fi
}%
\providecommand \natexlab [1]{#1}%
\providecommand \enquote  [1]{``#1''}%
\providecommand \bibnamefont  [1]{#1}%
\providecommand \bibfnamefont [1]{#1}%
\providecommand \citenamefont [1]{#1}%
\providecommand \href@noop [0]{\@secondoftwo}%
\providecommand \href [0]{\begingroup \@sanitize@url \@href}%
\providecommand \@href[1]{\@@startlink{#1}\@@href}%
\providecommand \@@href[1]{\endgroup#1\@@endlink}%
\providecommand \@sanitize@url [0]{\catcode `\\12\catcode `\$12\catcode
  `\&12\catcode `\#12\catcode `\^12\catcode `\_12\catcode `\%12\relax}%
\providecommand \@@startlink[1]{}%
\providecommand \@@endlink[0]{}%
\providecommand \url  [0]{\begingroup\@sanitize@url \@url }%
\providecommand \@url [1]{\endgroup\@href {#1}{\urlprefix }}%
\providecommand \urlprefix  [0]{URL }%
\providecommand \Eprint [0]{\href }%
\providecommand \doibase [0]{http://dx.doi.org/}%
\providecommand \selectlanguage [0]{\@gobble}%
\providecommand \bibinfo  [0]{\@secondoftwo}%
\providecommand \bibfield  [0]{\@secondoftwo}%
\providecommand \translation [1]{[#1]}%
\providecommand \BibitemOpen [0]{}%
\providecommand \bibitemStop [0]{}%
\providecommand \bibitemNoStop [0]{.\EOS\space}%
\providecommand \EOS [0]{\spacefactor3000\relax}%
\providecommand \BibitemShut  [1]{\csname bibitem#1\endcsname}%
\let\auto@bib@innerbib\@empty
\bibitem [{\citenamefont {Giesen}\ \emph {et~al.}(2015)\citenamefont {Giesen},
  \citenamefont {Boudaud}, \citenamefont {Genolini}, \citenamefont {Poulin},
  \citenamefont {Cirelli} \emph {et~al.}}]{Giesen:2015ufa}%
  \BibitemOpen
  \bibfield  {author} {\bibinfo {author} {\bibfnamefont {G.}~\bibnamefont
  {Giesen}}, \bibinfo {author} {\bibfnamefont {M.}~\bibnamefont {Boudaud}},
  \bibinfo {author} {\bibfnamefont {Y.}~\bibnamefont {Genolini}}, \bibinfo
  {author} {\bibfnamefont {V.}~\bibnamefont {Poulin}}, \bibinfo {author}
  {\bibfnamefont {M.}~\bibnamefont {Cirelli}},  \emph {et~al.},\ }\href@noop {}
  {\  (\bibinfo {year} {2015})},\ \Eprint {http://arxiv.org/abs/1504.04276}
  {arXiv:1504.04276 [astro-ph.HE]} \BibitemShut {NoStop}%
\bibitem [{\citenamefont {Jin}\ \emph {et~al.}(2015)\citenamefont {Jin},
  \citenamefont {Wu},\ and\ \citenamefont {Zhou}}]{Jin:2015mka}%
  \BibitemOpen
  \bibfield  {author} {\bibinfo {author} {\bibfnamefont {H.-B.}\ \bibnamefont
  {Jin}}, \bibinfo {author} {\bibfnamefont {Y.-L.}\ \bibnamefont {Wu}}, \ and\
  \bibinfo {author} {\bibfnamefont {Y.-F.}\ \bibnamefont {Zhou}}\ }(\bibinfo
  {year} {2015})\ \Eprint {http://arxiv.org/abs/1508.06844} {arXiv:1508.06844
  [hep-ph]} \BibitemShut {NoStop}%
\bibitem [{\citenamefont {Chen}\ \emph {et~al.}(2015)\citenamefont {Chen},
  \citenamefont {Chiang},\ and\ \citenamefont {Nomura}}]{Chen:2015cqa}%
  \BibitemOpen
  \bibfield  {author} {\bibinfo {author} {\bibfnamefont {C.-H.}\ \bibnamefont
  {Chen}}, \bibinfo {author} {\bibfnamefont {C.-W.}\ \bibnamefont {Chiang}}, \
  and\ \bibinfo {author} {\bibfnamefont {T.}~\bibnamefont {Nomura}},\
  }\href@noop {} {\  (\bibinfo {year} {2015})},\ \Eprint
  {http://arxiv.org/abs/1504.07848} {arXiv:1504.07848 [hep-ph]} \BibitemShut
  {NoStop}%
\bibitem [{\citenamefont {Ibe}\ \emph {et~al.}(2015)\citenamefont {Ibe},
  \citenamefont {Matsumoto}, \citenamefont {Shirai},\ and\ \citenamefont
  {Yanagida}}]{Ibe:2015tma}%
  \BibitemOpen
  \bibfield  {author} {\bibinfo {author} {\bibfnamefont {M.}~\bibnamefont
  {Ibe}}, \bibinfo {author} {\bibfnamefont {S.}~\bibnamefont {Matsumoto}},
  \bibinfo {author} {\bibfnamefont {S.}~\bibnamefont {Shirai}}, \ and\ \bibinfo
  {author} {\bibfnamefont {T.~T.}\ \bibnamefont {Yanagida}},\ }\href@noop {} {\
   (\bibinfo {year} {2015})},\ \Eprint {http://arxiv.org/abs/1504.05554}
  {arXiv:1504.05554 [hep-ph]} \BibitemShut {NoStop}%
\bibitem [{\citenamefont {Hamaguchi}\ \emph {et~al.}(2015)\citenamefont
  {Hamaguchi}, \citenamefont {Moroi},\ and\ \citenamefont
  {Nakayama}}]{Hamaguchi:2015wga}%
  \BibitemOpen
  \bibfield  {author} {\bibinfo {author} {\bibfnamefont {K.}~\bibnamefont
  {Hamaguchi}}, \bibinfo {author} {\bibfnamefont {T.}~\bibnamefont {Moroi}}, \
  and\ \bibinfo {author} {\bibfnamefont {K.}~\bibnamefont {Nakayama}},\
  }\href@noop {} {\  (\bibinfo {year} {2015})},\ \Eprint
  {http://arxiv.org/abs/1504.05937} {arXiv:1504.05937 [hep-ph]} \BibitemShut
  {NoStop}%
\bibitem [{\citenamefont {Kohri}\ \emph {et~al.}(2015)\citenamefont {Kohri},
  \citenamefont {Ioka}, \citenamefont {Fujita},\ and\ \citenamefont
  {Yamazaki}}]{Kohri:2015mga}%
  \BibitemOpen
  \bibfield  {author} {\bibinfo {author} {\bibfnamefont {K.}~\bibnamefont
  {Kohri}}, \bibinfo {author} {\bibfnamefont {K.}~\bibnamefont {Ioka}},
  \bibinfo {author} {\bibfnamefont {Y.}~\bibnamefont {Fujita}}, \ and\ \bibinfo
  {author} {\bibfnamefont {R.}~\bibnamefont {Yamazaki}},\ }\href@noop {} {\
  (\bibinfo {year} {2015})},\ \Eprint {http://arxiv.org/abs/1505.01236}
  {arXiv:1505.01236 [astro-ph.HE]} \BibitemShut {NoStop}%
\bibitem [{\citenamefont {Kappl}\ \emph {et~al.}(2015)\citenamefont {Kappl},
  \citenamefont {Reinert},\ and\ \citenamefont {Winkler}}]{Kappl:2015bqa}%
  \BibitemOpen
  \bibfield  {author} {\bibinfo {author} {\bibfnamefont {R.}~\bibnamefont
  {Kappl}}, \bibinfo {author} {\bibfnamefont {A.}~\bibnamefont {Reinert}}, \
  and\ \bibinfo {author} {\bibfnamefont {M.~W.}\ \bibnamefont {Winkler}},\
  }\href@noop {} {\  (\bibinfo {year} {2015})},\ \Eprint
  {http://arxiv.org/abs/1506.04145} {arXiv:1506.04145 [astro-ph.HE]}
  \BibitemShut {NoStop}%
\bibitem [{\citenamefont {Lu}\ and\ \citenamefont {Zong}(2015)}]{Lu:2015pta}%
  \BibitemOpen
  \bibfield  {author} {\bibinfo {author} {\bibfnamefont {B.-Q.}\ \bibnamefont
  {Lu}}\ and\ \bibinfo {author} {\bibfnamefont {H.-S.}\ \bibnamefont {Zong}},\
  }\href@noop {} {\  (\bibinfo {year} {2015})},\ \Eprint
  {http://arxiv.org/abs/1510.04032} {arXiv:1510.04032 [astro-ph.HE]}
  \BibitemShut {NoStop}%
\bibitem [{\citenamefont {Lin}\ \emph {et~al.}(2015)\citenamefont {Lin},
  \citenamefont {Bi}, \citenamefont {Yin},\ and\ \citenamefont
  {Yu}}]{Lin:2015taa}%
  \BibitemOpen
  \bibfield  {author} {\bibinfo {author} {\bibfnamefont {S.-J.}\ \bibnamefont
  {Lin}}, \bibinfo {author} {\bibfnamefont {X.-J.}\ \bibnamefont {Bi}},
  \bibinfo {author} {\bibfnamefont {P.-F.}\ \bibnamefont {Yin}}, \ and\
  \bibinfo {author} {\bibfnamefont {Z.-H.}\ \bibnamefont {Yu}},\ }\href@noop {}
  {\  (\bibinfo {year} {2015})},\ \Eprint {http://arxiv.org/abs/1504.07230}
  {arXiv:1504.07230 [hep-ph]} \BibitemShut {NoStop}%
\bibitem [{\citenamefont {Cui}\ \emph {et~al.}(2016)\citenamefont {Cui},
  \citenamefont {Yuan}, \citenamefont {Tsai},\ and\ \citenamefont
  {Fan}}]{Cui:2016ppb}%
  \BibitemOpen
  \bibfield  {author} {\bibinfo {author} {\bibfnamefont {M.-Y.}\ \bibnamefont
  {Cui}}, \bibinfo {author} {\bibfnamefont {Q.}~\bibnamefont {Yuan}}, \bibinfo
  {author} {\bibfnamefont {Y.-L.~S.}\ \bibnamefont {Tsai}}, \ and\ \bibinfo
  {author} {\bibfnamefont {Y.-Z.}\ \bibnamefont {Fan}},\ }\href@noop {} {\
  (\bibinfo {year} {2016})},\ \Eprint {http://arxiv.org/abs/1610.03840}
  {arXiv:1610.03840 [astro-ph.HE]} \BibitemShut {NoStop}%
\bibitem [{\citenamefont {Cuoco}\ \emph {et~al.}(2017)\citenamefont {Cuoco},
  \citenamefont {Krämer},\ and\ \citenamefont {Korsmeier}}]{Cuoco:2016eej}%
  \BibitemOpen
  \bibfield  {author} {\bibinfo {author} {\bibfnamefont {A.}~\bibnamefont
  {Cuoco}}, \bibinfo {author} {\bibfnamefont {M.}~\bibnamefont {Krämer}}, \
  and\ \bibinfo {author} {\bibfnamefont {M.}~\bibnamefont {Korsmeier}},\ }\href
  {\doibase 10.1103/PhysRevLett.118.191102} {\bibfield  {journal} {\bibinfo
  {journal} {Phys. Rev. Lett.}\ }\textbf {\bibinfo {volume} {118}},\ \bibinfo
  {pages} {191102} (\bibinfo {year} {2017})},\ \Eprint
  {http://arxiv.org/abs/1610.03071} {arXiv:1610.03071 [astro-ph.HE]}
  \BibitemShut {NoStop}%
\bibitem [{\citenamefont {Feng}\ \emph {et~al.}(2016)\citenamefont {Feng},
  \citenamefont {Tomassetti},\ and\ \citenamefont {Oliva}}]{Feng:2016loc}%
  \BibitemOpen
  \bibfield  {author} {\bibinfo {author} {\bibfnamefont {J.}~\bibnamefont
  {Feng}}, \bibinfo {author} {\bibfnamefont {N.}~\bibnamefont {Tomassetti}}, \
  and\ \bibinfo {author} {\bibfnamefont {A.}~\bibnamefont {Oliva}},\
  }\href@noop {} {\  (\bibinfo {year} {2016})},\ \Eprint
  {http://arxiv.org/abs/1610.06182} {arXiv:1610.06182 [astro-ph.HE]}
  \BibitemShut {NoStop}%
\bibitem [{\citenamefont {Huang}\ \emph {et~al.}(2017)\citenamefont {Huang},
  \citenamefont {Wei}, \citenamefont {Wu}, \citenamefont {Zhang},\ and\
  \citenamefont {Zhou}}]{Huang:2016tfo}%
  \BibitemOpen
  \bibfield  {author} {\bibinfo {author} {\bibfnamefont {X.-J.}\ \bibnamefont
  {Huang}}, \bibinfo {author} {\bibfnamefont {C.-C.}\ \bibnamefont {Wei}},
  \bibinfo {author} {\bibfnamefont {Y.-L.}\ \bibnamefont {Wu}}, \bibinfo
  {author} {\bibfnamefont {W.-H.}\ \bibnamefont {Zhang}}, \ and\ \bibinfo
  {author} {\bibfnamefont {Y.-F.}\ \bibnamefont {Zhou}},\ }\href {\doibase
  10.1103/PhysRevD.95.063021} {\bibfield  {journal} {\bibinfo  {journal} {Phys.
  Rev.}\ }\textbf {\bibinfo {volume} {D95}},\ \bibinfo {pages} {063021}
  (\bibinfo {year} {2017})},\ \Eprint {http://arxiv.org/abs/1611.01983}
  {arXiv:1611.01983 [hep-ph]} \BibitemShut {NoStop}%
\bibitem [{\citenamefont {Lin}\ \emph {et~al.}(2016)\citenamefont {Lin},
  \citenamefont {Bi}, \citenamefont {Feng}, \citenamefont {Yin},\ and\
  \citenamefont {Yu}}]{Lin:2016ezz}%
  \BibitemOpen
  \bibfield  {author} {\bibinfo {author} {\bibfnamefont {S.-J.}\ \bibnamefont
  {Lin}}, \bibinfo {author} {\bibfnamefont {X.-J.}\ \bibnamefont {Bi}},
  \bibinfo {author} {\bibfnamefont {J.}~\bibnamefont {Feng}}, \bibinfo {author}
  {\bibfnamefont {P.-F.}\ \bibnamefont {Yin}}, \ and\ \bibinfo {author}
  {\bibfnamefont {Z.-H.}\ \bibnamefont {Yu}},\ }\href@noop {} {\  (\bibinfo
  {year} {2016})},\ \Eprint {http://arxiv.org/abs/1612.04001} {arXiv:1612.04001
  [astro-ph.HE]} \BibitemShut {NoStop}%
\bibitem [{\citenamefont {Cui}\ \emph {et~al.}(2018{\natexlab{a}})\citenamefont
  {Cui}, \citenamefont {Pan}, \citenamefont {Yuan}, \citenamefont {Fan},\ and\
  \citenamefont {Zong}}]{Cui:2018klo}%
  \BibitemOpen
  \bibfield  {author} {\bibinfo {author} {\bibfnamefont {M.-Y.}\ \bibnamefont
  {Cui}}, \bibinfo {author} {\bibfnamefont {X.}~\bibnamefont {Pan}}, \bibinfo
  {author} {\bibfnamefont {Q.}~\bibnamefont {Yuan}}, \bibinfo {author}
  {\bibfnamefont {Y.-Z.}\ \bibnamefont {Fan}}, \ and\ \bibinfo {author}
  {\bibfnamefont {H.-S.}\ \bibnamefont {Zong}},\ }\href {\doibase
  10.1088/1475-7516/2018/06/024} {\bibfield  {journal} {\bibinfo  {journal}
  {JCAP}\ }\textbf {\bibinfo {volume} {1806}},\ \bibinfo {pages} {024}
  (\bibinfo {year} {2018}{\natexlab{a}})},\ \Eprint
  {http://arxiv.org/abs/1803.02163} {arXiv:1803.02163 [astro-ph.HE]}
  \BibitemShut {NoStop}%
\bibitem [{\citenamefont {Cui}\ \emph {et~al.}(2018{\natexlab{b}})\citenamefont
  {Cui}, \citenamefont {Huang}, \citenamefont {Tsai},\ and\ \citenamefont
  {Yuan}}]{Cui:2018nlm}%
  \BibitemOpen
  \bibfield  {author} {\bibinfo {author} {\bibfnamefont {M.-Y.}\ \bibnamefont
  {Cui}}, \bibinfo {author} {\bibfnamefont {W.-C.}\ \bibnamefont {Huang}},
  \bibinfo {author} {\bibfnamefont {Y.-L.~S.}\ \bibnamefont {Tsai}}, \ and\
  \bibinfo {author} {\bibfnamefont {Q.}~\bibnamefont {Yuan}},\ }\href {\doibase
  10.1088/1475-7516/2018/11/039} {\bibfield  {journal} {\bibinfo  {journal}
  {JCAP}\ }\textbf {\bibinfo {volume} {1811}},\ \bibinfo {pages} {039}
  (\bibinfo {year} {2018}{\natexlab{b}})},\ \Eprint
  {http://arxiv.org/abs/1805.11590} {arXiv:1805.11590 [hep-ph]} \BibitemShut
  {NoStop}%
\bibitem [{\citenamefont {Cuoco}\ \emph {et~al.}(2019)\citenamefont {Cuoco},
  \citenamefont {Heisig}, \citenamefont {Klamt}, \citenamefont {Korsmeier},\
  and\ \citenamefont {Krämer}}]{Cuoco:2019kuu}%
  \BibitemOpen
  \bibfield  {author} {\bibinfo {author} {\bibfnamefont {A.}~\bibnamefont
  {Cuoco}}, \bibinfo {author} {\bibfnamefont {J.}~\bibnamefont {Heisig}},
  \bibinfo {author} {\bibfnamefont {L.}~\bibnamefont {Klamt}}, \bibinfo
  {author} {\bibfnamefont {M.}~\bibnamefont {Korsmeier}}, \ and\ \bibinfo
  {author} {\bibfnamefont {M.}~\bibnamefont {Krämer}},\ }\href@noop {} {\
  (\bibinfo {year} {2019})},\ \Eprint {http://arxiv.org/abs/1903.01472}
  {arXiv:1903.01472 [astro-ph.HE]} \BibitemShut {NoStop}%
\bibitem [{\citenamefont {Cholis}\ \emph {et~al.}(2019)\citenamefont {Cholis},
  \citenamefont {Linden},\ and\ \citenamefont {Hooper}}]{Cholis:2019ejx}%
  \BibitemOpen
  \bibfield  {author} {\bibinfo {author} {\bibfnamefont {I.}~\bibnamefont
  {Cholis}}, \bibinfo {author} {\bibfnamefont {T.}~\bibnamefont {Linden}}, \
  and\ \bibinfo {author} {\bibfnamefont {D.}~\bibnamefont {Hooper}},\
  }\href@noop {} {\  (\bibinfo {year} {2019})},\ \Eprint
  {http://arxiv.org/abs/1903.02549} {arXiv:1903.02549 [astro-ph.HE]}
  \BibitemShut {NoStop}%
\bibitem [{\citenamefont {Kalmykov}\ \emph {et~al.}(1997)\citenamefont
  {Kalmykov}, \citenamefont {Ostapchenko},\ and\ \citenamefont
  {Pavlov}}]{Kalmykov:1997te}%
  \BibitemOpen
  \bibfield  {author} {\bibinfo {author} {\bibfnamefont {N.~N.}\ \bibnamefont
  {Kalmykov}}, \bibinfo {author} {\bibfnamefont {S.~S.}\ \bibnamefont
  {Ostapchenko}}, \ and\ \bibinfo {author} {\bibfnamefont {A.~I.}\ \bibnamefont
  {Pavlov}},\ }\bibfield  {booktitle} {\emph {\bibinfo {booktitle}
  {Proceedings, 9th International Symposium on Very High Energy Cosmic Ray
  Interactions (ISVHECRI 1996)}},\ }\href {\doibase
  10.1016/S0920-5632(96)00846-8} {\bibfield  {journal} {\bibinfo  {journal}
  {Nucl. Phys. Proc. Suppl.}\ }\textbf {\bibinfo {volume} {52}},\ \bibinfo
  {pages} {17} (\bibinfo {year} {1997})}\BibitemShut {NoStop}%
\bibitem [{\citenamefont
  {Ostapchenko}(2006{\natexlab{a}})}]{Ostapchenko:2004ss}%
  \BibitemOpen
  \bibfield  {author} {\bibinfo {author} {\bibfnamefont {S.}~\bibnamefont
  {Ostapchenko}},\ }\bibfield  {booktitle} {\emph {\bibinfo {booktitle}
  {Proceedings, 13th International Symposium on Very High-Energy Cosmic Ray
  Interactions (ISVHECRI 2004)}},\ }\href {\doibase
  10.1016/j.nuclphysbps.2005.07.026} {\bibfield  {journal} {\bibinfo  {journal}
  {Nucl. Phys. Proc. Suppl.}\ }\textbf {\bibinfo {volume} {151}},\ \bibinfo
  {pages} {143} (\bibinfo {year} {2006}{\natexlab{a}})},\ \Eprint
  {http://arxiv.org/abs/hep-ph/0412332} {arXiv:hep-ph/0412332 [hep-ph]}
  \BibitemShut {NoStop}%
\bibitem [{\citenamefont {Engel}\ \emph {et~al.}(1992)\citenamefont {Engel},
  \citenamefont {Gaisser}, \citenamefont {Stanev},\ and\ \citenamefont
  {Lipari}}]{Engel:1992vf}%
  \BibitemOpen
  \bibfield  {author} {\bibinfo {author} {\bibfnamefont {J.}~\bibnamefont
  {Engel}}, \bibinfo {author} {\bibfnamefont {T.~K.}\ \bibnamefont {Gaisser}},
  \bibinfo {author} {\bibfnamefont {T.}~\bibnamefont {Stanev}}, \ and\ \bibinfo
  {author} {\bibfnamefont {P.}~\bibnamefont {Lipari}},\ }\href {\doibase
  10.1103/PhysRevD.46.5013} {\bibfield  {journal} {\bibinfo  {journal} {Phys.
  Rev.}\ }\textbf {\bibinfo {volume} {D46}},\ \bibinfo {pages} {5013} (\bibinfo
  {year} {1992})}\BibitemShut {NoStop}%
\bibitem [{\citenamefont {Engel}(1995)}]{Engel:1994vs}%
  \BibitemOpen
  \bibfield  {author} {\bibinfo {author} {\bibfnamefont {R.}~\bibnamefont
  {Engel}},\ }\href {\doibase 10.1007/BF01496594} {\bibfield  {journal}
  {\bibinfo  {journal} {Z. Phys.}\ }\textbf {\bibinfo {volume} {C66}},\
  \bibinfo {pages} {203} (\bibinfo {year} {1995})}\BibitemShut {NoStop}%
\bibitem [{\citenamefont {Bopp}\ \emph {et~al.}(2008)\citenamefont {Bopp},
  \citenamefont {Ranft}, \citenamefont {Engel},\ and\ \citenamefont
  {Roesler}}]{Bopp:2005cr}%
  \BibitemOpen
  \bibfield  {author} {\bibinfo {author} {\bibfnamefont {F.~W.}\ \bibnamefont
  {Bopp}}, \bibinfo {author} {\bibfnamefont {J.}~\bibnamefont {Ranft}},
  \bibinfo {author} {\bibfnamefont {R.}~\bibnamefont {Engel}}, \ and\ \bibinfo
  {author} {\bibfnamefont {S.}~\bibnamefont {Roesler}},\ }\href {\doibase
  10.1103/PhysRevC.77.014904} {\bibfield  {journal} {\bibinfo  {journal} {Phys.
  Rev.}\ }\textbf {\bibinfo {volume} {C77}},\ \bibinfo {pages} {014904}
  (\bibinfo {year} {2008})},\ \Eprint {http://arxiv.org/abs/hep-ph/0505035}
  {arXiv:hep-ph/0505035 [hep-ph]} \BibitemShut {NoStop}%
\bibitem [{\citenamefont {Werner}\ \emph {et~al.}(2006)\citenamefont {Werner},
  \citenamefont {Liu},\ and\ \citenamefont {Pierog}}]{Werner:2005jf}%
  \BibitemOpen
  \bibfield  {author} {\bibinfo {author} {\bibfnamefont {K.}~\bibnamefont
  {Werner}}, \bibinfo {author} {\bibfnamefont {F.-M.}\ \bibnamefont {Liu}}, \
  and\ \bibinfo {author} {\bibfnamefont {T.}~\bibnamefont {Pierog}},\ }\href
  {\doibase 10.1103/PhysRevC.74.044902} {\bibfield  {journal} {\bibinfo
  {journal} {Phys. Rev.}\ }\textbf {\bibinfo {volume} {C74}},\ \bibinfo {pages}
  {044902} (\bibinfo {year} {2006})},\ \Eprint
  {http://arxiv.org/abs/hep-ph/0506232} {arXiv:hep-ph/0506232 [hep-ph]}
  \BibitemShut {NoStop}%
\bibitem [{\citenamefont {Pierog}\ \emph {et~al.}(2015)\citenamefont {Pierog},
  \citenamefont {Karpenko}, \citenamefont {Katzy}, \citenamefont {Yatsenko},\
  and\ \citenamefont {Werner}}]{Pierog:2013ria}%
  \BibitemOpen
  \bibfield  {author} {\bibinfo {author} {\bibfnamefont {T.}~\bibnamefont
  {Pierog}}, \bibinfo {author} {\bibfnamefont {I.}~\bibnamefont {Karpenko}},
  \bibinfo {author} {\bibfnamefont {J.~M.}\ \bibnamefont {Katzy}}, \bibinfo
  {author} {\bibfnamefont {E.}~\bibnamefont {Yatsenko}}, \ and\ \bibinfo
  {author} {\bibfnamefont {K.}~\bibnamefont {Werner}},\ }\href {\doibase
  10.1103/PhysRevC.92.034906} {\bibfield  {journal} {\bibinfo  {journal} {Phys.
  Rev.}\ }\textbf {\bibinfo {volume} {C92}},\ \bibinfo {pages} {034906}
  (\bibinfo {year} {2015})},\ \Eprint {http://arxiv.org/abs/1306.0121}
  {arXiv:1306.0121 [hep-ph]} \BibitemShut {NoStop}%
\bibitem [{\citenamefont {Tan}\ and\ \citenamefont
  {Ng}(1983)}]{1983JPhG....9.1289T}%
  \BibitemOpen
  \bibfield  {author} {\bibinfo {author} {\bibfnamefont {L.~C.}\ \bibnamefont
  {Tan}}\ and\ \bibinfo {author} {\bibfnamefont {L.~K.}\ \bibnamefont {Ng}},\
  }\href {\doibase 10.1088/0305-4616/9/10/015} {\bibfield  {journal} {\bibinfo
  {journal} {Journal of Physics G Nuclear Physics}\ }\textbf {\bibinfo {volume}
  {9}},\ \bibinfo {pages} {1289} (\bibinfo {year} {1983})}\BibitemShut
  {NoStop}%
\bibitem [{\citenamefont {di~Mauro}\ \emph {et~al.}(2014)\citenamefont
  {di~Mauro}, \citenamefont {Donato}, \citenamefont {Goudelis},\ and\
  \citenamefont {Serpico}}]{diMauro:2014zea}%
  \BibitemOpen
  \bibfield  {author} {\bibinfo {author} {\bibfnamefont {M.}~\bibnamefont
  {di~Mauro}}, \bibinfo {author} {\bibfnamefont {F.}~\bibnamefont {Donato}},
  \bibinfo {author} {\bibfnamefont {A.}~\bibnamefont {Goudelis}}, \ and\
  \bibinfo {author} {\bibfnamefont {P.~D.}\ \bibnamefont {Serpico}},\ }\href
  {\doibase 10.1103/PhysRevD.90.085017} {\bibfield  {journal} {\bibinfo
  {journal} {Phys.Rev.}\ }\textbf {\bibinfo {volume} {D90}},\ \bibinfo {pages}
  {085017} (\bibinfo {year} {2014})},\ \Eprint {http://arxiv.org/abs/1408.0288}
  {arXiv:1408.0288 [hep-ph]} \BibitemShut {NoStop}%
\bibitem [{\citenamefont {Winkler}(2017)}]{Winkler:2017xor}%
  \BibitemOpen
  \bibfield  {author} {\bibinfo {author} {\bibfnamefont {M.~W.}\ \bibnamefont
  {Winkler}},\ }\href@noop {} {\  (\bibinfo {year} {2017})},\ \Eprint
  {http://arxiv.org/abs/1701.04866} {arXiv:1701.04866 [hep-ph]} \BibitemShut
  {NoStop}%
\bibitem [{\citenamefont {Strong}\ and\ \citenamefont
  {Moskalenko}(1998)}]{Strong:1998pw}%
  \BibitemOpen
  \bibfield  {author} {\bibinfo {author} {\bibfnamefont {A.}~\bibnamefont
  {Strong}}\ and\ \bibinfo {author} {\bibfnamefont {I.}~\bibnamefont
  {Moskalenko}},\ }\href {\doibase 10.1086/306470} {\bibfield  {journal}
  {\bibinfo  {journal} {Astrophys.J.}\ }\textbf {\bibinfo {volume} {509}},\
  \bibinfo {pages} {212} (\bibinfo {year} {1998})},\ \Eprint
  {http://arxiv.org/abs/astro-ph/9807150} {arXiv:astro-ph/9807150 [astro-ph]}
  \BibitemShut {NoStop}%
\bibitem [{\citenamefont {Gaisser}(1990)}]{1990cup..book.....G}%
  \BibitemOpen
  \bibfield  {author} {\bibinfo {author} {\bibfnamefont {T.~K.}\ \bibnamefont
  {Gaisser}},\ }\href {http://adsabs.harvard.edu/abs/1990cup..book.....G}
  {\emph {\bibinfo {title} {Cambridge and New York, Cambridge University Press,
  1990, 292 p.}}}\ (\bibinfo {year} {1990})\BibitemShut {NoStop}%
\bibitem [{\citenamefont {Di~Bernardo}\ \emph {et~al.}(2010)\citenamefont
  {Di~Bernardo}, \citenamefont {Evoli}, \citenamefont {Gaggero}, \citenamefont
  {Grasso},\ and\ \citenamefont {Maccione}}]{DiBernardo:2009ku}%
  \BibitemOpen
  \bibfield  {author} {\bibinfo {author} {\bibfnamefont {G.}~\bibnamefont
  {Di~Bernardo}}, \bibinfo {author} {\bibfnamefont {C.}~\bibnamefont {Evoli}},
  \bibinfo {author} {\bibfnamefont {D.}~\bibnamefont {Gaggero}}, \bibinfo
  {author} {\bibfnamefont {D.}~\bibnamefont {Grasso}}, \ and\ \bibinfo {author}
  {\bibfnamefont {L.}~\bibnamefont {Maccione}},\ }\href {\doibase
  10.1016/j.astropartphys.2010.08.006} {\bibfield  {journal} {\bibinfo
  {journal} {Astropart. Phys.}\ }\textbf {\bibinfo {volume} {34}},\ \bibinfo
  {pages} {274} (\bibinfo {year} {2010})},\ \Eprint
  {http://arxiv.org/abs/0909.4548} {arXiv:0909.4548 [astro-ph.HE]} \BibitemShut
  {NoStop}%
\bibitem [{\citenamefont {Maurin}\ \emph {et~al.}(2001)\citenamefont {Maurin},
  \citenamefont {Donato}, \citenamefont {Taillet},\ and\ \citenamefont
  {Salati}}]{Maurin:2001sj}%
  \BibitemOpen
  \bibfield  {author} {\bibinfo {author} {\bibfnamefont {D.}~\bibnamefont
  {Maurin}}, \bibinfo {author} {\bibfnamefont {F.}~\bibnamefont {Donato}},
  \bibinfo {author} {\bibfnamefont {R.}~\bibnamefont {Taillet}}, \ and\
  \bibinfo {author} {\bibfnamefont {P.}~\bibnamefont {Salati}},\ }\href
  {\doibase 10.1086/321496} {\bibfield  {journal} {\bibinfo  {journal}
  {Astrophys. J.}\ }\textbf {\bibinfo {volume} {555}},\ \bibinfo {pages} {585}
  (\bibinfo {year} {2001})},\ \Eprint {http://arxiv.org/abs/astro-ph/0101231}
  {arXiv:astro-ph/0101231 [astro-ph]} \BibitemShut {NoStop}%
\bibitem [{\citenamefont {Yuan}\ \emph {et~al.}(2017)\citenamefont {Yuan},
  \citenamefont {Lin}, \citenamefont {Fang},\ and\ \citenamefont
  {Bi}}]{Yuan:2017ozr}%
  \BibitemOpen
  \bibfield  {author} {\bibinfo {author} {\bibfnamefont {Q.}~\bibnamefont
  {Yuan}}, \bibinfo {author} {\bibfnamefont {S.-J.}\ \bibnamefont {Lin}},
  \bibinfo {author} {\bibfnamefont {K.}~\bibnamefont {Fang}}, \ and\ \bibinfo
  {author} {\bibfnamefont {X.-J.}\ \bibnamefont {Bi}},\ }\href {\doibase
  10.1103/PhysRevD.95.083007} {\bibfield  {journal} {\bibinfo  {journal} {Phys.
  Rev.}\ }\textbf {\bibinfo {volume} {D95}},\ \bibinfo {pages} {083007}
  (\bibinfo {year} {2017})},\ \Eprint {http://arxiv.org/abs/1701.06149}
  {arXiv:1701.06149 [astro-ph.HE]} \BibitemShut {NoStop}%
\bibitem [{\citenamefont {Moskalenko}\ and\ \citenamefont
  {Strong}(1998)}]{Moskalenko:1997gh}%
  \BibitemOpen
  \bibfield  {author} {\bibinfo {author} {\bibfnamefont {I.~V.}\ \bibnamefont
  {Moskalenko}}\ and\ \bibinfo {author} {\bibfnamefont {A.~W.}\ \bibnamefont
  {Strong}},\ }\href {\doibase 10.1086/305152} {\bibfield  {journal} {\bibinfo
  {journal} {Astrophys. J.}\ }\textbf {\bibinfo {volume} {493}},\ \bibinfo
  {pages} {694} (\bibinfo {year} {1998})},\ \Eprint
  {http://arxiv.org/abs/astro-ph/9710124} {arXiv:astro-ph/9710124 [astro-ph]}
  \BibitemShut {NoStop}%
\bibitem [{\citenamefont {Evoli}\ \emph {et~al.}(2017)\citenamefont {Evoli},
  \citenamefont {Gaggero}, \citenamefont {Vittino}, \citenamefont
  {Di~Bernardo}, \citenamefont {Di~Mauro}, \citenamefont {Ligorini},
  \citenamefont {Ullio},\ and\ \citenamefont {Grasso}}]{Evoli:2016xgn}%
  \BibitemOpen
  \bibfield  {author} {\bibinfo {author} {\bibfnamefont {C.}~\bibnamefont
  {Evoli}}, \bibinfo {author} {\bibfnamefont {D.}~\bibnamefont {Gaggero}},
  \bibinfo {author} {\bibfnamefont {A.}~\bibnamefont {Vittino}}, \bibinfo
  {author} {\bibfnamefont {G.}~\bibnamefont {Di~Bernardo}}, \bibinfo {author}
  {\bibfnamefont {M.}~\bibnamefont {Di~Mauro}}, \bibinfo {author}
  {\bibfnamefont {A.}~\bibnamefont {Ligorini}}, \bibinfo {author}
  {\bibfnamefont {P.}~\bibnamefont {Ullio}}, \ and\ \bibinfo {author}
  {\bibfnamefont {D.}~\bibnamefont {Grasso}},\ }\href {\doibase
  10.1088/1475-7516/2017/02/015} {\bibfield  {journal} {\bibinfo  {journal}
  {JCAP}\ }\textbf {\bibinfo {volume} {1702}},\ \bibinfo {pages} {015}
  (\bibinfo {year} {2017})},\ \Eprint {http://arxiv.org/abs/1607.07886}
  {arXiv:1607.07886 [astro-ph.HE]} \BibitemShut {NoStop}%
\bibitem [{\citenamefont {Strong}\ \emph {et~al.}(2007)\citenamefont {Strong},
  \citenamefont {Moskalenko},\ and\ \citenamefont {Ptuskin}}]{Strong:2007nh}%
  \BibitemOpen
  \bibfield  {author} {\bibinfo {author} {\bibfnamefont {A.~W.}\ \bibnamefont
  {Strong}}, \bibinfo {author} {\bibfnamefont {I.~V.}\ \bibnamefont
  {Moskalenko}}, \ and\ \bibinfo {author} {\bibfnamefont {V.~S.}\ \bibnamefont
  {Ptuskin}},\ }\href {\doibase 10.1146/annurev.nucl.57.090506.123011}
  {\bibfield  {journal} {\bibinfo  {journal} {Ann.Rev.Nucl.Part.Sci.}\ }\textbf
  {\bibinfo {volume} {57}},\ \bibinfo {pages} {285} (\bibinfo {year} {2007})},\
  \Eprint {http://arxiv.org/abs/astro-ph/0701517} {arXiv:astro-ph/0701517
  [astro-ph]} \BibitemShut {NoStop}%
\bibitem [{\citenamefont {Putze}\ \emph {et~al.}(2010)\citenamefont {Putze},
  \citenamefont {Derome},\ and\ \citenamefont {Maurin}}]{Putze:2010zn}%
  \BibitemOpen
  \bibfield  {author} {\bibinfo {author} {\bibfnamefont {A.}~\bibnamefont
  {Putze}}, \bibinfo {author} {\bibfnamefont {L.}~\bibnamefont {Derome}}, \
  and\ \bibinfo {author} {\bibfnamefont {D.}~\bibnamefont {Maurin}},\ }\href
  {\doibase 10.1051/0004-6361/201014010} {\bibfield  {journal} {\bibinfo
  {journal} {Astron. Astrophys.}\ }\textbf {\bibinfo {volume} {516}},\ \bibinfo
  {pages} {A66} (\bibinfo {year} {2010})},\ \Eprint
  {http://arxiv.org/abs/1001.0551} {arXiv:1001.0551 [astro-ph.HE]} \BibitemShut
  {NoStop}%
\bibitem [{\citenamefont {Trotta}\ \emph {et~al.}(2011)\citenamefont {Trotta},
  \citenamefont {Johannesson}, \citenamefont {Moskalenko}, \citenamefont
  {Porter}, \citenamefont {de~Austri} \emph {et~al.}}]{Trotta:2010mx}%
  \BibitemOpen
  \bibfield  {author} {\bibinfo {author} {\bibfnamefont {R.}~\bibnamefont
  {Trotta}}, \bibinfo {author} {\bibfnamefont {G.}~\bibnamefont {Johannesson}},
  \bibinfo {author} {\bibfnamefont {I.}~\bibnamefont {Moskalenko}}, \bibinfo
  {author} {\bibfnamefont {T.}~\bibnamefont {Porter}}, \bibinfo {author}
  {\bibfnamefont {R.~R.}\ \bibnamefont {de~Austri}},  \emph {et~al.},\ }\href
  {\doibase 10.1088/0004-637X/729/2/106} {\bibfield  {journal} {\bibinfo
  {journal} {Astrophys.J.}\ }\textbf {\bibinfo {volume} {729}},\ \bibinfo
  {pages} {106} (\bibinfo {year} {2011})},\ \Eprint
  {http://arxiv.org/abs/1011.0037} {arXiv:1011.0037 [astro-ph.HE]} \BibitemShut
  {NoStop}%
\bibitem [{\citenamefont {Aguilar}\ \emph
  {et~al.}(2016{\natexlab{a}})\citenamefont {Aguilar} \emph
  {et~al.}}]{Aguilar:2016vqr}%
  \BibitemOpen
  \bibfield  {author} {\bibinfo {author} {\bibfnamefont {M.}~\bibnamefont
  {Aguilar}} \emph {et~al.} (\bibinfo {collaboration} {AMS}),\ }\href {\doibase
  10.1103/PhysRevLett.117.231102} {\bibfield  {journal} {\bibinfo  {journal}
  {Phys. Rev. Lett.}\ }\textbf {\bibinfo {volume} {117}},\ \bibinfo {pages}
  {231102} (\bibinfo {year} {2016}{\natexlab{a}})}\BibitemShut {NoStop}%
\bibitem [{\citenamefont {Maurin}\ \emph {et~al.}(2010)\citenamefont {Maurin},
  \citenamefont {Putze},\ and\ \citenamefont {Derome}}]{Maurin:2010zp}%
  \BibitemOpen
  \bibfield  {author} {\bibinfo {author} {\bibfnamefont {D.}~\bibnamefont
  {Maurin}}, \bibinfo {author} {\bibfnamefont {A.}~\bibnamefont {Putze}}, \
  and\ \bibinfo {author} {\bibfnamefont {L.}~\bibnamefont {Derome}},\ }\href
  {\doibase 10.1051/0004-6361/201014011} {\bibfield  {journal} {\bibinfo
  {journal} {Astron.Astrophys.}\ }\textbf {\bibinfo {volume} {516}},\ \bibinfo
  {pages} {A67} (\bibinfo {year} {2010})},\ \Eprint
  {http://arxiv.org/abs/1001.0553} {arXiv:1001.0553 [astro-ph.HE]} \BibitemShut
  {NoStop}%
\bibitem [{\citenamefont {Di~Bernardo}\ \emph {et~al.}(2011)\citenamefont
  {Di~Bernardo}, \citenamefont {Evoli}, \citenamefont {Gaggero}, \citenamefont
  {Grasso}, \citenamefont {Maccione},\ and\ \citenamefont
  {Mazziotta}}]{DiBernardo:2010is}%
  \BibitemOpen
  \bibfield  {author} {\bibinfo {author} {\bibfnamefont {G.}~\bibnamefont
  {Di~Bernardo}}, \bibinfo {author} {\bibfnamefont {C.}~\bibnamefont {Evoli}},
  \bibinfo {author} {\bibfnamefont {D.}~\bibnamefont {Gaggero}}, \bibinfo
  {author} {\bibfnamefont {D.}~\bibnamefont {Grasso}}, \bibinfo {author}
  {\bibfnamefont {L.}~\bibnamefont {Maccione}}, \ and\ \bibinfo {author}
  {\bibfnamefont {M.~N.}\ \bibnamefont {Mazziotta}},\ }\href {\doibase
  10.1016/j.astropartphys.2010.11.005} {\bibfield  {journal} {\bibinfo
  {journal} {Astropart. Phys.}\ }\textbf {\bibinfo {volume} {34}},\ \bibinfo
  {pages} {528} (\bibinfo {year} {2011})},\ \Eprint
  {http://arxiv.org/abs/1010.0174} {arXiv:1010.0174 [astro-ph.HE]} \BibitemShut
  {NoStop}%
\bibitem [{\citenamefont {Seo}\ and\ \citenamefont
  {Ptuskin}(1994)}]{1994ApJ...431..705S}%
  \BibitemOpen
  \bibfield  {author} {\bibinfo {author} {\bibfnamefont {E.~S.}\ \bibnamefont
  {Seo}}\ and\ \bibinfo {author} {\bibfnamefont {V.~S.}\ \bibnamefont
  {Ptuskin}},\ }\href {\doibase 10.1086/174520} {\bibfield  {journal} {\bibinfo
   {journal} {\apj}\ }\textbf {\bibinfo {volume} {431}},\ \bibinfo {pages}
  {705} (\bibinfo {year} {1994})}\BibitemShut {NoStop}%
\bibitem [{\citenamefont {Parker}(1965)}]{1965P&SS...13....9P}%
  \BibitemOpen
  \bibfield  {author} {\bibinfo {author} {\bibfnamefont {E.~N.}\ \bibnamefont
  {Parker}},\ }\href {\doibase 10.1016/0032-0633(65)90131-5} {\bibfield
  {journal} {\bibinfo  {journal} {\planss}\ }\textbf {\bibinfo {volume} {13}},\
  \bibinfo {pages} {9} (\bibinfo {year} {1965})}\BibitemShut {NoStop}%
\bibitem [{\citenamefont {Gleeson}\ and\ \citenamefont
  {Axford}(1968)}]{Gleeson:1968zza}%
  \BibitemOpen
  \bibfield  {author} {\bibinfo {author} {\bibfnamefont {L.~J.}\ \bibnamefont
  {Gleeson}}\ and\ \bibinfo {author} {\bibfnamefont {W.~I.}\ \bibnamefont
  {Axford}},\ }\href {\doibase 10.1086/149822} {\bibfield  {journal} {\bibinfo
  {journal} {Astrophys. J.}\ }\textbf {\bibinfo {volume} {154}},\ \bibinfo
  {pages} {1011} (\bibinfo {year} {1968})}\BibitemShut {NoStop}%
\bibitem [{\citenamefont {Potgieter}(2014)}]{Potgieter:2014pka}%
  \BibitemOpen
  \bibfield  {author} {\bibinfo {author} {\bibfnamefont {M.~S.}\ \bibnamefont
  {Potgieter}},\ }\bibfield  {booktitle} {\emph {\bibinfo {booktitle}
  {Proceedings, 39th COSPAR Scientific Assembly. Cosmic Ray Origins: Viktor
  Hess Centennial Anniversary: Mysore, India, July 14-22, 2012}},\ }\href
  {\doibase 10.1016/j.asr.2013.04.015} {\bibfield  {journal} {\bibinfo
  {journal} {Adv. Space Res.}\ }\textbf {\bibinfo {volume} {53}},\ \bibinfo
  {pages} {1415} (\bibinfo {year} {2014})}\BibitemShut {NoStop}%
\bibitem [{\citenamefont {Corcella}\ \emph {et~al.}(2001)\citenamefont
  {Corcella}, \citenamefont {Knowles}, \citenamefont {Marchesini},
  \citenamefont {Moretti}, \citenamefont {Odagiri}, \citenamefont {Richardson},
  \citenamefont {Seymour},\ and\ \citenamefont {Webber}}]{Corcella:2000bw}%
  \BibitemOpen
  \bibfield  {author} {\bibinfo {author} {\bibfnamefont {G.}~\bibnamefont
  {Corcella}}, \bibinfo {author} {\bibfnamefont {I.~G.}\ \bibnamefont
  {Knowles}}, \bibinfo {author} {\bibfnamefont {G.}~\bibnamefont {Marchesini}},
  \bibinfo {author} {\bibfnamefont {S.}~\bibnamefont {Moretti}}, \bibinfo
  {author} {\bibfnamefont {K.}~\bibnamefont {Odagiri}}, \bibinfo {author}
  {\bibfnamefont {P.}~\bibnamefont {Richardson}}, \bibinfo {author}
  {\bibfnamefont {M.~H.}\ \bibnamefont {Seymour}}, \ and\ \bibinfo {author}
  {\bibfnamefont {B.~R.}\ \bibnamefont {Webber}},\ }\href {\doibase
  10.1088/1126-6708/2001/01/010} {\bibfield  {journal} {\bibinfo  {journal}
  {JHEP}\ }\textbf {\bibinfo {volume} {01}},\ \bibinfo {pages} {010} (\bibinfo
  {year} {2001})},\ \Eprint {http://arxiv.org/abs/hep-ph/0011363}
  {arXiv:hep-ph/0011363 [hep-ph]} \BibitemShut {NoStop}%
\bibitem [{\citenamefont {Sjostrand}\ \emph {et~al.}(2006)\citenamefont
  {Sjostrand}, \citenamefont {Mrenna},\ and\ \citenamefont
  {Skands}}]{Sjostrand:2006za}%
  \BibitemOpen
  \bibfield  {author} {\bibinfo {author} {\bibfnamefont {T.}~\bibnamefont
  {Sjostrand}}, \bibinfo {author} {\bibfnamefont {S.}~\bibnamefont {Mrenna}}, \
  and\ \bibinfo {author} {\bibfnamefont {P.~Z.}\ \bibnamefont {Skands}},\
  }\href {\doibase 10.1088/1126-6708/2006/05/026} {\bibfield  {journal}
  {\bibinfo  {journal} {JHEP}\ }\textbf {\bibinfo {volume} {05}},\ \bibinfo
  {pages} {026} (\bibinfo {year} {2006})},\ \Eprint
  {http://arxiv.org/abs/hep-ph/0603175} {arXiv:hep-ph/0603175 [hep-ph]}
  \BibitemShut {NoStop}%
\bibitem [{\citenamefont {Gleisberg}\ \emph {et~al.}(2009)\citenamefont
  {Gleisberg}, \citenamefont {Hoeche}, \citenamefont {Krauss}, \citenamefont
  {Schonherr}, \citenamefont {Schumann}, \citenamefont {Siegert},\ and\
  \citenamefont {Winter}}]{Gleisberg:2008ta}%
  \BibitemOpen
  \bibfield  {author} {\bibinfo {author} {\bibfnamefont {T.}~\bibnamefont
  {Gleisberg}}, \bibinfo {author} {\bibfnamefont {S.}~\bibnamefont {Hoeche}},
  \bibinfo {author} {\bibfnamefont {F.}~\bibnamefont {Krauss}}, \bibinfo
  {author} {\bibfnamefont {M.}~\bibnamefont {Schonherr}}, \bibinfo {author}
  {\bibfnamefont {S.}~\bibnamefont {Schumann}}, \bibinfo {author}
  {\bibfnamefont {F.}~\bibnamefont {Siegert}}, \ and\ \bibinfo {author}
  {\bibfnamefont {J.}~\bibnamefont {Winter}},\ }\href {\doibase
  10.1088/1126-6708/2009/02/007} {\bibfield  {journal} {\bibinfo  {journal}
  {JHEP}\ }\textbf {\bibinfo {volume} {02}},\ \bibinfo {pages} {007} (\bibinfo
  {year} {2009})},\ \Eprint {http://arxiv.org/abs/0811.4622} {arXiv:0811.4622
  [hep-ph]} \BibitemShut {NoStop}%
\bibitem [{\citenamefont
  {Ostapchenko}(2006{\natexlab{b}})}]{Ostapchenko:2005nj}%
  \BibitemOpen
  \bibfield  {author} {\bibinfo {author} {\bibfnamefont {S.}~\bibnamefont
  {Ostapchenko}},\ }\href {\doibase 10.1103/PhysRevD.74.014026} {\bibfield
  {journal} {\bibinfo  {journal} {Phys. Rev.}\ }\textbf {\bibinfo {volume}
  {D74}},\ \bibinfo {pages} {014026} (\bibinfo {year} {2006}{\natexlab{b}})},\
  \Eprint {http://arxiv.org/abs/hep-ph/0505259} {arXiv:hep-ph/0505259 [hep-ph]}
  \BibitemShut {NoStop}%
\bibitem [{\citenamefont {Fletcher}\ \emph {et~al.}(1994)\citenamefont
  {Fletcher}, \citenamefont {Gaisser}, \citenamefont {Lipari},\ and\
  \citenamefont {Stanev}}]{Fletcher:1994bd}%
  \BibitemOpen
  \bibfield  {author} {\bibinfo {author} {\bibfnamefont {R.~S.}\ \bibnamefont
  {Fletcher}}, \bibinfo {author} {\bibfnamefont {T.~K.}\ \bibnamefont
  {Gaisser}}, \bibinfo {author} {\bibfnamefont {P.}~\bibnamefont {Lipari}}, \
  and\ \bibinfo {author} {\bibfnamefont {T.}~\bibnamefont {Stanev}},\ }\href
  {\doibase 10.1103/PhysRevD.50.5710} {\bibfield  {journal} {\bibinfo
  {journal} {Phys. Rev.}\ }\textbf {\bibinfo {volume} {D50}},\ \bibinfo {pages}
  {5710} (\bibinfo {year} {1994})}\BibitemShut {NoStop}%
\bibitem [{\citenamefont {Ahn}\ \emph {et~al.}(2009)\citenamefont {Ahn},
  \citenamefont {Engel}, \citenamefont {Gaisser}, \citenamefont {Lipari},\ and\
  \citenamefont {Stanev}}]{Ahn:2009wx}%
  \BibitemOpen
  \bibfield  {author} {\bibinfo {author} {\bibfnamefont {E.-J.}\ \bibnamefont
  {Ahn}}, \bibinfo {author} {\bibfnamefont {R.}~\bibnamefont {Engel}}, \bibinfo
  {author} {\bibfnamefont {T.~K.}\ \bibnamefont {Gaisser}}, \bibinfo {author}
  {\bibfnamefont {P.}~\bibnamefont {Lipari}}, \ and\ \bibinfo {author}
  {\bibfnamefont {T.}~\bibnamefont {Stanev}},\ }\href {\doibase
  10.1103/PhysRevD.80.094003} {\bibfield  {journal} {\bibinfo  {journal} {Phys.
  Rev.}\ }\textbf {\bibinfo {volume} {D80}},\ \bibinfo {pages} {094003}
  (\bibinfo {year} {2009})},\ \Eprint {http://arxiv.org/abs/0906.4113}
  {arXiv:0906.4113 [hep-ph]} \BibitemShut {NoStop}%
\bibitem [{\citenamefont {Engel}\ and\ \citenamefont
  {Ranft}(1996)}]{Engel:1995yda}%
  \BibitemOpen
  \bibfield  {author} {\bibinfo {author} {\bibfnamefont {R.}~\bibnamefont
  {Engel}}\ and\ \bibinfo {author} {\bibfnamefont {J.}~\bibnamefont {Ranft}},\
  }\href {\doibase 10.1103/PhysRevD.54.4244} {\bibfield  {journal} {\bibinfo
  {journal} {Phys. Rev.}\ }\textbf {\bibinfo {volume} {D54}},\ \bibinfo {pages}
  {4244} (\bibinfo {year} {1996})},\ \Eprint
  {http://arxiv.org/abs/hep-ph/9509373} {arXiv:hep-ph/9509373 [hep-ph]}
  \BibitemShut {NoStop}%
\bibitem [{\citenamefont {Engel}\ \emph {et~al.}(1995)\citenamefont {Engel},
  \citenamefont {Ranft},\ and\ \citenamefont {Roesler}}]{Engel:1995sb}%
  \BibitemOpen
  \bibfield  {author} {\bibinfo {author} {\bibfnamefont {R.}~\bibnamefont
  {Engel}}, \bibinfo {author} {\bibfnamefont {J.}~\bibnamefont {Ranft}}, \ and\
  \bibinfo {author} {\bibfnamefont {S.}~\bibnamefont {Roesler}},\ }\href
  {\doibase 10.1103/PhysRevD.52.1459} {\bibfield  {journal} {\bibinfo
  {journal} {Phys. Rev.}\ }\textbf {\bibinfo {volume} {D52}},\ \bibinfo {pages}
  {1459} (\bibinfo {year} {1995})},\ \Eprint
  {http://arxiv.org/abs/hep-ph/9502319} {arXiv:hep-ph/9502319 [hep-ph]}
  \BibitemShut {NoStop}%
\bibitem [{\citenamefont {Albrow}\ \emph {et~al.}(1973)\citenamefont {Albrow}
  \emph {et~al.}}]{Albrow:1973kj}%
  \BibitemOpen
  \bibfield  {author} {\bibinfo {author} {\bibfnamefont {M.~G.}\ \bibnamefont
  {Albrow}} \emph {et~al.} (\bibinfo {collaboration} {CHLM}),\ }\href {\doibase
  10.1016/0550-3213(73)90035-7} {\bibfield  {journal} {\bibinfo  {journal}
  {Nucl. Phys.}\ }\textbf {\bibinfo {volume} {B56}},\ \bibinfo {pages} {333}
  (\bibinfo {year} {1973})}\BibitemShut {NoStop}%
\bibitem [{\citenamefont {Anticic}\ \emph {et~al.}(2010)\citenamefont {Anticic}
  \emph {et~al.}}]{Anticic:2009wd}%
  \BibitemOpen
  \bibfield  {author} {\bibinfo {author} {\bibfnamefont {T.}~\bibnamefont
  {Anticic}} \emph {et~al.} (\bibinfo {collaboration} {NA49}),\ }\href
  {\doibase 10.1140/epjc/s10052-009-1172-2} {\bibfield  {journal} {\bibinfo
  {journal} {Eur. Phys. J.}\ }\textbf {\bibinfo {volume} {C65}},\ \bibinfo
  {pages} {9} (\bibinfo {year} {2010})},\ \Eprint
  {http://arxiv.org/abs/0904.2708} {arXiv:0904.2708 [hep-ex]} \BibitemShut
  {NoStop}%
\bibitem [{\citenamefont {Pierog}\ and\ \citenamefont {Ulrich}()}]{CRMCWeb}%
  \BibitemOpen
  \bibfield  {author} {\bibinfo {author} {\bibfnamefont {T.}~\bibnamefont
  {Pierog}}\ and\ \bibinfo {author} {\bibfnamefont {R.}~\bibnamefont
  {Ulrich}},\ }\href {https://web.ikp.kit.edu/rulrich/crmc.html} {\enquote
  {\bibinfo {title} {{CRMC(Cosmic Ray Monte Carlo package)}},}\ }\bibinfo
  {howpublished} {\url{https://web.ikp.kit.edu/rulrich/crmc.html}}\BibitemShut
  {NoStop}%
\bibitem [{\citenamefont {Pierog}\ and\ \citenamefont
  {Werner}(2009)}]{Pierog:2009zt}%
  \BibitemOpen
  \bibfield  {author} {\bibinfo {author} {\bibfnamefont {T.}~\bibnamefont
  {Pierog}}\ and\ \bibinfo {author} {\bibfnamefont {K.}~\bibnamefont
  {Werner}},\ }\bibfield  {booktitle} {\emph {\bibinfo {booktitle}
  {Proceedings, 15th International Symposium on Very High Energy Cosmic Ray
  Interactions (ISVHECRI 2008): Paris, France, September 1-6, 2008}},\ }\href
  {\doibase 10.1016/j.nuclphysbps.2009.09.017} {\bibfield  {journal} {\bibinfo
  {journal} {Nucl. Phys. Proc. Suppl.}\ }\textbf {\bibinfo {volume} {196}},\
  \bibinfo {pages} {102} (\bibinfo {year} {2009})},\ \Eprint
  {http://arxiv.org/abs/0905.1198} {arXiv:0905.1198 [hep-ph]} \BibitemShut
  {NoStop}%
\bibitem [{\citenamefont {Laszlo}\ \emph {et~al.}(2007)\citenamefont {Laszlo}
  \emph {et~al.}}]{Laszlo:2007ib}%
  \BibitemOpen
  \bibfield  {author} {\bibinfo {author} {\bibfnamefont {A.}~\bibnamefont
  {Laszlo}} \emph {et~al.} (\bibinfo {collaboration} {NA61}),\ }\bibfield
  {booktitle} {\emph {\bibinfo {booktitle} {Proceedings, 4th International
  Workshop on Critical point and onset of deconfinement (CPOD07): Darmstadt,
  Germany, July 9-13, 2007}},\ }\href {\doibase 10.22323/1.047.0054} {\bibfield
   {journal} {\bibinfo  {journal} {PoS}\ }\textbf {\bibinfo {volume}
  {CPOD07}},\ \bibinfo {pages} {054} (\bibinfo {year} {2007})},\ \Eprint
  {http://arxiv.org/abs/0709.1867} {arXiv:0709.1867 [nucl-ex]} \BibitemShut
  {NoStop}%
\bibitem [{\citenamefont {Kachelriess}\ \emph {et~al.}(2015)\citenamefont
  {Kachelriess}, \citenamefont {Moskalenko},\ and\ \citenamefont
  {Ostapchenko}}]{Kachelriess:2015wpa}%
  \BibitemOpen
  \bibfield  {author} {\bibinfo {author} {\bibfnamefont {M.}~\bibnamefont
  {Kachelriess}}, \bibinfo {author} {\bibfnamefont {I.~V.}\ \bibnamefont
  {Moskalenko}}, \ and\ \bibinfo {author} {\bibfnamefont {S.~S.}\ \bibnamefont
  {Ostapchenko}},\ }\href@noop {} {\  (\bibinfo {year} {2015})},\ \Eprint
  {http://arxiv.org/abs/1502.04158} {arXiv:1502.04158 [astro-ph.HE]}
  \BibitemShut {NoStop}%
\bibitem [{\citenamefont {Duperray}\ \emph {et~al.}(2003)\citenamefont
  {Duperray}, \citenamefont {Huang}, \citenamefont {Protasov},\ and\
  \citenamefont {Buenerd}}]{Duperray:2003bd}%
  \BibitemOpen
  \bibfield  {author} {\bibinfo {author} {\bibfnamefont {R.~P.}\ \bibnamefont
  {Duperray}}, \bibinfo {author} {\bibfnamefont {C.~Y.}\ \bibnamefont {Huang}},
  \bibinfo {author} {\bibfnamefont {K.~V.}\ \bibnamefont {Protasov}}, \ and\
  \bibinfo {author} {\bibfnamefont {M.}~\bibnamefont {Buenerd}},\ }\href
  {\doibase 10.1103/PhysRevD.68.094017} {\bibfield  {journal} {\bibinfo
  {journal} {Phys. Rev.}\ }\textbf {\bibinfo {volume} {D68}},\ \bibinfo {pages}
  {094017} (\bibinfo {year} {2003})},\ \Eprint
  {http://arxiv.org/abs/astro-ph/0305274} {arXiv:astro-ph/0305274 [astro-ph]}
  \BibitemShut {NoStop}%
\bibitem [{\citenamefont {Arsene}\ \emph {et~al.}(2007)\citenamefont {Arsene}
  \emph {et~al.}}]{Arsene:2007jd}%
  \BibitemOpen
  \bibfield  {author} {\bibinfo {author} {\bibfnamefont {I.}~\bibnamefont
  {Arsene}} \emph {et~al.} (\bibinfo {collaboration} {BRAHMS}),\ }\href
  {\doibase 10.1103/PhysRevLett.98.252001} {\bibfield  {journal} {\bibinfo
  {journal} {Phys. Rev. Lett.}\ }\textbf {\bibinfo {volume} {98}},\ \bibinfo
  {pages} {252001} (\bibinfo {year} {2007})},\ \Eprint
  {http://arxiv.org/abs/hep-ex/0701041} {arXiv:hep-ex/0701041 [hep-ex]}
  \BibitemShut {NoStop}%
\bibitem [{GAL(2012)}]{GALPROPTheoryWebSite}%
  \BibitemOpen
  \href {http://galprop.stanford.edu/code.php?option=theory} {\enquote
  {\bibinfo {title} {{GALPROP: Code - Theory}},}\ }\bibinfo {howpublished}
  {\url{http://galprop.stanford.edu/code.php?option=theory}} (\bibinfo {year}
  {2012}),\ \bibinfo {note} {accessed: 2012-09-05}\BibitemShut {NoStop}%
\bibitem [{\citenamefont {Cirelli}\ \emph {et~al.}(2011)\citenamefont
  {Cirelli}, \citenamefont {Corcella}, \citenamefont {Hektor}, \citenamefont
  {Hutsi}, \citenamefont {Kadastik} \emph {et~al.}}]{Cirelli:2010xx}%
  \BibitemOpen
  \bibfield  {author} {\bibinfo {author} {\bibfnamefont {M.}~\bibnamefont
  {Cirelli}}, \bibinfo {author} {\bibfnamefont {G.}~\bibnamefont {Corcella}},
  \bibinfo {author} {\bibfnamefont {A.}~\bibnamefont {Hektor}}, \bibinfo
  {author} {\bibfnamefont {G.}~\bibnamefont {Hutsi}}, \bibinfo {author}
  {\bibfnamefont {M.}~\bibnamefont {Kadastik}},  \emph {et~al.},\ }\href
  {\doibase 10.1088/1475-7516/2012/10/E01, 10.1088/1475-7516/2011/03/051}
  {\bibfield  {journal} {\bibinfo  {journal} {JCAP}\ }\textbf {\bibinfo
  {volume} {1103}},\ \bibinfo {pages} {051} (\bibinfo {year} {2011})},\ \Eprint
  {http://arxiv.org/abs/1012.4515} {arXiv:1012.4515 [hep-ph]} \BibitemShut
  {NoStop}%
\bibitem [{\citenamefont {Navarro}\ \emph {et~al.}(1997)\citenamefont
  {Navarro}, \citenamefont {Frenk},\ and\ \citenamefont
  {White}}]{Navarro:1996gj}%
  \BibitemOpen
  \bibfield  {author} {\bibinfo {author} {\bibfnamefont {J.~F.}\ \bibnamefont
  {Navarro}}, \bibinfo {author} {\bibfnamefont {C.~S.}\ \bibnamefont {Frenk}},
  \ and\ \bibinfo {author} {\bibfnamefont {S.~D.}\ \bibnamefont {White}},\
  }\href {\doibase 10.1086/304888} {\bibfield  {journal} {\bibinfo  {journal}
  {Astrophys.J.}\ }\textbf {\bibinfo {volume} {490}},\ \bibinfo {pages} {493}
  (\bibinfo {year} {1997})},\ \Eprint {http://arxiv.org/abs/astro-ph/9611107}
  {arXiv:astro-ph/9611107 [astro-ph]} \BibitemShut {NoStop}%
\bibitem [{\citenamefont {Nesti}\ and\ \citenamefont
  {Salucci}(2013)}]{Nesti:2013uwa}%
  \BibitemOpen
  \bibfield  {author} {\bibinfo {author} {\bibfnamefont {F.}~\bibnamefont
  {Nesti}}\ and\ \bibinfo {author} {\bibfnamefont {P.}~\bibnamefont
  {Salucci}},\ }\href {\doibase 10.1088/1475-7516/2013/07/016} {\bibfield
  {journal} {\bibinfo  {journal} {JCAP}\ }\textbf {\bibinfo {volume} {1307}},\
  \bibinfo {pages} {016} (\bibinfo {year} {2013})},\ \Eprint
  {http://arxiv.org/abs/1304.5127} {arXiv:1304.5127 [astro-ph.GA]} \BibitemShut
  {NoStop}%
\bibitem [{\citenamefont {Benito}\ \emph {et~al.}(2019)\citenamefont {Benito},
  \citenamefont {Cuoco},\ and\ \citenamefont {Iocco}}]{Benito:2019ngh}%
  \BibitemOpen
  \bibfield  {author} {\bibinfo {author} {\bibfnamefont {M.}~\bibnamefont
  {Benito}}, \bibinfo {author} {\bibfnamefont {A.}~\bibnamefont {Cuoco}}, \
  and\ \bibinfo {author} {\bibfnamefont {F.}~\bibnamefont {Iocco}},\
  }\href@noop {} {\  (\bibinfo {year} {2019})},\ \Eprint
  {http://arxiv.org/abs/1901.02460} {arXiv:1901.02460 [astro-ph.GA]}
  \BibitemShut {NoStop}%
\bibitem [{\citenamefont {Karukes}\ \emph {et~al.}(2019)\citenamefont
  {Karukes}, \citenamefont {Benito}, \citenamefont {Iocco}, \citenamefont
  {Trotta},\ and\ \citenamefont {Geringer-Sameth}}]{Karukes:2019jxv}%
  \BibitemOpen
  \bibfield  {author} {\bibinfo {author} {\bibfnamefont {E.~V.}\ \bibnamefont
  {Karukes}}, \bibinfo {author} {\bibfnamefont {M.}~\bibnamefont {Benito}},
  \bibinfo {author} {\bibfnamefont {F.}~\bibnamefont {Iocco}}, \bibinfo
  {author} {\bibfnamefont {R.}~\bibnamefont {Trotta}}, \ and\ \bibinfo {author}
  {\bibfnamefont {A.}~\bibnamefont {Geringer-Sameth}},\ }\href@noop {} {\
  (\bibinfo {year} {2019})},\ \Eprint {http://arxiv.org/abs/1901.02463}
  {arXiv:1901.02463 [astro-ph.GA]} \BibitemShut {NoStop}%
\bibitem [{\citenamefont {Cortes}(1995)}]{Cortes1995}%
  \BibitemOpen
  \bibfield  {author} {\bibinfo {author} {\bibfnamefont {V.}~\bibnamefont
  {Cortes}, \bibfnamefont {Corinna~Vapnik}},\ }\href {\doibase
  10.1007/BF00994018} {\bibfield  {journal} {\bibinfo  {journal} {Machine
  Learning}\ }\textbf {\bibinfo {volume} {20}},\ \bibinfo {pages} {273}
  (\bibinfo {year} {1995})}\BibitemShut {NoStop}%
\bibitem [{\citenamefont {Altman}(1992)}]{Altman:1992ait}%
  \BibitemOpen
  \bibfield  {author} {\bibinfo {author} {\bibfnamefont {N.~S.}\ \bibnamefont
  {Altman}},\ }\href {\doibase 10.1080/00031305.1992.10475879} {\bibfield
  {journal} {\bibinfo  {journal} {The American Statistician}\ }\textbf
  {\bibinfo {volume} {46}},\ \bibinfo {pages} {175} (\bibinfo {year} {1992})},\
  \Eprint
  {http://arxiv.org/abs/https://www.tandfonline.com/doi/pdf/10.1080/00031305.1992.10475879}
  {https://www.tandfonline.com/doi/pdf/10.1080/00031305.1992.10475879}
  \BibitemShut {NoStop}%
\bibitem [{\citenamefont {Kami\'nski}(2018)}]{Kamiński2018}%
  \BibitemOpen
  \bibfield  {author} {\bibinfo {author} {\bibfnamefont {M.~S.~P.}\
  \bibnamefont {Kami\'nski}, \bibfnamefont {Bogumi\l~Jakubczyk}},\ }\href
  {\doibase 10.1007/s10100-017-0479-6} {\bibfield  {journal} {\bibinfo
  {journal} {Central European Journal of Operations Research}\ }\textbf
  {\bibinfo {volume} {26}},\ \bibinfo {pages} {135} (\bibinfo {year}
  {2018})}\BibitemShut {NoStop}%
\bibitem [{\citenamefont {Ho}(1998)}]{Ho:1998trs}%
  \BibitemOpen
  \bibfield  {author} {\bibinfo {author} {\bibfnamefont {T.~K.}\ \bibnamefont
  {Ho}},\ }\href {\doibase 10.1109/34.709601} {\bibfield  {journal} {\bibinfo
  {journal} {IEEE Transactions on Pattern Analysis and Machine Intelligence}\
  }\textbf {\bibinfo {volume} {20}},\ \bibinfo {pages} {832} (\bibinfo {year}
  {1998})}\BibitemShut {NoStop}%
\bibitem [{\citenamefont {McCulloch}(1943)}]{McCulloch1943}%
  \BibitemOpen
  \bibfield  {author} {\bibinfo {author} {\bibfnamefont {W.}~\bibnamefont
  {McCulloch}, \bibfnamefont {Warren S.~Pitts}},\ }\href {\doibase
  10.1007/BF02478259} {\bibfield  {journal} {\bibinfo  {journal} {The bulletin
  of mathematical biophysics}\ }\textbf {\bibinfo {volume} {5}},\ \bibinfo
  {pages} {115} (\bibinfo {year} {1943})}\BibitemShut {NoStop}%
\bibitem [{\citenamefont {Chang}\ and\ \citenamefont
  {Lin}(2011)}]{Chang:2011:LLS:1961189.1961199}%
  \BibitemOpen
  \bibfield  {author} {\bibinfo {author} {\bibfnamefont {C.-C.}\ \bibnamefont
  {Chang}}\ and\ \bibinfo {author} {\bibfnamefont {C.-J.}\ \bibnamefont
  {Lin}},\ }\href {\doibase 10.1145/1961189.1961199} {\bibfield  {journal}
  {\bibinfo  {journal} {ACM Trans. Intell. Syst. Technol.}\ }\textbf {\bibinfo
  {volume} {2}},\ \bibinfo {pages} {27:1} (\bibinfo {year} {2011})}\BibitemShut
  {NoStop}%
\bibitem [{\citenamefont
  {Spyromitros-Xioufis}(2016)}]{Spyromitros-Xioufis2016}%
  \BibitemOpen
  \bibfield  {author} {\bibinfo {author} {\bibfnamefont {G.~G. W. V.~I.}\
  \bibnamefont {Spyromitros-Xioufis}, \bibfnamefont {Eleftherios~Tsoumakas}},\
  }\href {\doibase 10.1007/s10994-016-5546-z} {\bibfield  {journal} {\bibinfo
  {journal} {Machine Learning}\ }\textbf {\bibinfo {volume} {104}},\ \bibinfo
  {pages} {55} (\bibinfo {year} {2016})}\BibitemShut {NoStop}%
\bibitem [{\citenamefont {Aguilar}(2015)}]{Aguilar:2015ooa}%
  \BibitemOpen
  \bibfield  {author} {\bibinfo {author} {\bibfnamefont {M.}~\bibnamefont
  {Aguilar}} (\bibinfo {collaboration} {AMS}),\ }\href {\doibase
  10.1103/PhysRevLett.114.171103} {\bibfield  {journal} {\bibinfo  {journal}
  {Phys.Rev.Lett.}\ }\textbf {\bibinfo {volume} {114}},\ \bibinfo {pages}
  {171103} (\bibinfo {year} {2015})}\BibitemShut {NoStop}%
\bibitem [{\citenamefont {Aguilar}\ \emph
  {et~al.}(2016{\natexlab{b}})\citenamefont {Aguilar} \emph
  {et~al.}}]{Aguilar:2016kjl}%
  \BibitemOpen
  \bibfield  {author} {\bibinfo {author} {\bibfnamefont {M.}~\bibnamefont
  {Aguilar}} \emph {et~al.} (\bibinfo {collaboration} {AMS}),\ }\href {\doibase
  10.1103/PhysRevLett.117.091103} {\bibfield  {journal} {\bibinfo  {journal}
  {Phys. Rev. Lett.}\ }\textbf {\bibinfo {volume} {117}},\ \bibinfo {pages}
  {091103} (\bibinfo {year} {2016}{\natexlab{b}})}\BibitemShut {NoStop}%
\bibitem [{NLO()}]{NLOPTWebSite}%
  \BibitemOpen
  \href {http://ab-initio.mit.edu/nlopt} {\enquote {\bibinfo {title} {{The
  NLopt nonlinear-optimization package}},}\ }\bibinfo {howpublished}
  {\url{http://ab-initio.mit.edu/nlopt}}\BibitemShut {NoStop}%
\bibitem [{\citenamefont {Birgin}\ and\ \citenamefont
  {Martínez}(2008)}]{Birgin:2008iuc}%
  \BibitemOpen
  \bibfield  {author} {\bibinfo {author} {\bibfnamefont {E.}~\bibnamefont
  {Birgin}}\ and\ \bibinfo {author} {\bibfnamefont {J.}~\bibnamefont
  {Martínez}},\ }\href {https://doi.org/10.1080/10556780701577730 = {
  https://doi.org/10.1080/10556780701577730 }, Doi =
  {10.1080/10556780701577730},} {\bibfield  {journal} {\bibinfo  {journal}
  {Optimization Methods and Software}\ }\textbf {\bibinfo {volume} {23}},\
  \bibinfo {pages} {177} (\bibinfo {year} {2008})}\BibitemShut {NoStop}%
\bibitem [{\citenamefont {Griest}\ and\ \citenamefont
  {Kamionkowski}(1990)}]{Griest:1989wd}%
  \BibitemOpen
  \bibfield  {author} {\bibinfo {author} {\bibfnamefont {K.}~\bibnamefont
  {Griest}}\ and\ \bibinfo {author} {\bibfnamefont {M.}~\bibnamefont
  {Kamionkowski}},\ }\href {\doibase 10.1103/PhysRevLett.64.615} {\bibfield
  {journal} {\bibinfo  {journal} {Phys. Rev. Lett.}\ }\textbf {\bibinfo
  {volume} {64}},\ \bibinfo {pages} {615} (\bibinfo {year} {1990})}\BibitemShut
  {NoStop}%
\bibitem [{\citenamefont {Ahnen}\ \emph {et~al.}(2016)\citenamefont {Ahnen}
  \emph {et~al.}}]{Ahnen:2016qkx}%
  \BibitemOpen
  \bibfield  {author} {\bibinfo {author} {\bibfnamefont {M.~L.}\ \bibnamefont
  {Ahnen}} \emph {et~al.} (\bibinfo {collaboration} {Fermi-LAT, MAGIC}),\
  }\href {\doibase 10.1088/1475-7516/2016/02/039} {\bibfield  {journal}
  {\bibinfo  {journal} {JCAP}\ }\textbf {\bibinfo {volume} {1602}},\ \bibinfo
  {pages} {039} (\bibinfo {year} {2016})},\ \Eprint
  {http://arxiv.org/abs/1601.06590} {arXiv:1601.06590 [astro-ph.HE]}
  \BibitemShut {NoStop}%
\bibitem [{\citenamefont {Xiang}\ \emph {et~al.}(2017)\citenamefont {Xiang},
  \citenamefont {Bi}, \citenamefont {Lin},\ and\ \citenamefont
  {Yin}}]{Xiang:2017jou}%
  \BibitemOpen
  \bibfield  {author} {\bibinfo {author} {\bibfnamefont {Q.-F.}\ \bibnamefont
  {Xiang}}, \bibinfo {author} {\bibfnamefont {X.-J.}\ \bibnamefont {Bi}},
  \bibinfo {author} {\bibfnamefont {S.-J.}\ \bibnamefont {Lin}}, \ and\
  \bibinfo {author} {\bibfnamefont {P.-F.}\ \bibnamefont {Yin}},\ }\href
  {\doibase 10.1016/j.physletb.2017.09.003} {\bibfield  {journal} {\bibinfo
  {journal} {Phys. Lett.}\ }\textbf {\bibinfo {volume} {B773}},\ \bibinfo
  {pages} {448} (\bibinfo {year} {2017})},\ \Eprint
  {http://arxiv.org/abs/1707.09313} {arXiv:1707.09313 [astro-ph.HE]}
  \BibitemShut {NoStop}%
\bibitem [{\citenamefont {Moskalenko}\ \emph {et~al.}(2003)\citenamefont
  {Moskalenko}, \citenamefont {Strong}, \citenamefont {Mashnik},\ and\
  \citenamefont {Ormes}}]{Moskalenko:2002yx}%
  \BibitemOpen
  \bibfield  {author} {\bibinfo {author} {\bibfnamefont {I.~V.}\ \bibnamefont
  {Moskalenko}}, \bibinfo {author} {\bibfnamefont {A.}~\bibnamefont {Strong}},
  \bibinfo {author} {\bibfnamefont {S.}~\bibnamefont {Mashnik}}, \ and\
  \bibinfo {author} {\bibfnamefont {J.}~\bibnamefont {Ormes}},\ }\href
  {\doibase 10.1086/367697} {\bibfield  {journal} {\bibinfo  {journal}
  {Astrophys.J.}\ }\textbf {\bibinfo {volume} {586}},\ \bibinfo {pages} {1050}
  (\bibinfo {year} {2003})},\ \Eprint {http://arxiv.org/abs/astro-ph/0210480}
  {arXiv:astro-ph/0210480 [astro-ph]} \BibitemShut {NoStop}%
\bibitem [{\citenamefont {Hooper}\ \emph {et~al.}(2015)\citenamefont {Hooper},
  \citenamefont {Linden},\ and\ \citenamefont {Mertsch}}]{Hooper:2014ysa}%
  \BibitemOpen
  \bibfield  {author} {\bibinfo {author} {\bibfnamefont {D.}~\bibnamefont
  {Hooper}}, \bibinfo {author} {\bibfnamefont {T.}~\bibnamefont {Linden}}, \
  and\ \bibinfo {author} {\bibfnamefont {P.}~\bibnamefont {Mertsch}},\ }\href
  {\doibase 10.1088/1475-7516/2015/03/021} {\bibfield  {journal} {\bibinfo
  {journal} {JCAP}\ }\textbf {\bibinfo {volume} {1503}},\ \bibinfo {pages}
  {021} (\bibinfo {year} {2015})},\ \Eprint {http://arxiv.org/abs/1410.1527}
  {arXiv:1410.1527 [astro-ph.HE]} \BibitemShut {NoStop}%
\bibitem [{\citenamefont {Evoli}\ \emph {et~al.}(2012)\citenamefont {Evoli},
  \citenamefont {Cholis}, \citenamefont {Grasso}, \citenamefont {Maccione},\
  and\ \citenamefont {Ullio}}]{Evoli:2011id}%
  \BibitemOpen
  \bibfield  {author} {\bibinfo {author} {\bibfnamefont {C.}~\bibnamefont
  {Evoli}}, \bibinfo {author} {\bibfnamefont {I.}~\bibnamefont {Cholis}},
  \bibinfo {author} {\bibfnamefont {D.}~\bibnamefont {Grasso}}, \bibinfo
  {author} {\bibfnamefont {L.}~\bibnamefont {Maccione}}, \ and\ \bibinfo
  {author} {\bibfnamefont {P.}~\bibnamefont {Ullio}},\ }\href {\doibase
  10.1103/PhysRevD.85.123511} {\bibfield  {journal} {\bibinfo  {journal} {Phys.
  Rev.}\ }\textbf {\bibinfo {volume} {D85}},\ \bibinfo {pages} {123511}
  (\bibinfo {year} {2012})},\ \Eprint {http://arxiv.org/abs/1108.0664}
  {arXiv:1108.0664 [astro-ph.HE]} \BibitemShut {NoStop}%
\bibitem [{\citenamefont {Maccione}(2013)}]{Maccione:2012cu}%
  \BibitemOpen
  \bibfield  {author} {\bibinfo {author} {\bibfnamefont {L.}~\bibnamefont
  {Maccione}},\ }\href {\doibase 10.1103/PhysRevLett.110.081101} {\bibfield
  {journal} {\bibinfo  {journal} {Phys.Rev.Lett.}\ }\textbf {\bibinfo {volume}
  {110}},\ \bibinfo {pages} {081101} (\bibinfo {year} {2013})},\ \Eprint
  {http://arxiv.org/abs/1211.6905} {arXiv:1211.6905 [astro-ph.HE]} \BibitemShut
  {NoStop}%
\bibitem [{\citenamefont {Kappl}(2016)}]{Kappl:2015hxv}%
  \BibitemOpen
  \bibfield  {author} {\bibinfo {author} {\bibfnamefont {R.}~\bibnamefont
  {Kappl}},\ }\href {\doibase 10.1016/j.cpc.2016.05.025} {\bibfield  {journal}
  {\bibinfo  {journal} {Comput. Phys. Commun.}\ }\textbf {\bibinfo {volume}
  {207}},\ \bibinfo {pages} {386} (\bibinfo {year} {2016})},\ \Eprint
  {http://arxiv.org/abs/1511.07875} {arXiv:1511.07875 [astro-ph.SR]}
  \BibitemShut {NoStop}%
\bibitem [{\citenamefont {Aaij}\ \emph {et~al.}(2018)\citenamefont {Aaij} \emph
  {et~al.}}]{Aaij:2018svt}%
  \BibitemOpen
  \bibfield  {author} {\bibinfo {author} {\bibfnamefont {R.}~\bibnamefont
  {Aaij}} \emph {et~al.} (\bibinfo {collaboration} {LHCb}),\ }\href {\doibase
  10.1103/PhysRevLett.121.222001} {\bibfield  {journal} {\bibinfo  {journal}
  {Phys. Rev. Lett.}\ }\textbf {\bibinfo {volume} {121}},\ \bibinfo {pages}
  {222001} (\bibinfo {year} {2018})},\ \Eprint
  {http://arxiv.org/abs/1808.06127} {arXiv:1808.06127 [hep-ex]} \BibitemShut
  {NoStop}%
\bibitem [{\citenamefont {Korsmeier}\ \emph {et~al.}(2018)\citenamefont
  {Korsmeier}, \citenamefont {Donato},\ and\ \citenamefont
  {Di~Mauro}}]{Korsmeier:2018gcy}%
  \BibitemOpen
  \bibfield  {author} {\bibinfo {author} {\bibfnamefont {M.}~\bibnamefont
  {Korsmeier}}, \bibinfo {author} {\bibfnamefont {F.}~\bibnamefont {Donato}}, \
  and\ \bibinfo {author} {\bibfnamefont {M.}~\bibnamefont {Di~Mauro}},\ }\href
  {\doibase 10.1103/PhysRevD.97.103019} {\bibfield  {journal} {\bibinfo
  {journal} {Phys. Rev.}\ }\textbf {\bibinfo {volume} {D97}},\ \bibinfo {pages}
  {103019} (\bibinfo {year} {2018})},\ \Eprint
  {http://arxiv.org/abs/1802.03030} {arXiv:1802.03030 [astro-ph.HE]}
  \BibitemShut {NoStop}%
\bibitem [{\citenamefont {Drucker}\ \emph {et~al.}(1996)\citenamefont
  {Drucker}, \citenamefont {Burges}, \citenamefont {Kaufman}, \citenamefont
  {Smola},\ and\ \citenamefont {Vapnik}}]{Drucker:1996svr}%
  \BibitemOpen
  \bibfield  {author} {\bibinfo {author} {\bibfnamefont {H.}~\bibnamefont
  {Drucker}}, \bibinfo {author} {\bibfnamefont {C.~J.~C.}\ \bibnamefont
  {Burges}}, \bibinfo {author} {\bibfnamefont {L.}~\bibnamefont {Kaufman}},
  \bibinfo {author} {\bibfnamefont {A.~J.}\ \bibnamefont {Smola}}, \ and\
  \bibinfo {author} {\bibfnamefont {V.~N.}\ \bibnamefont {Vapnik}},\
  }\href@noop {} {\bibfield  {journal} {\bibinfo  {journal} {Advances in Neural
  Information Processing Systems}\ }\textbf {\bibinfo {volume} {9}} (\bibinfo
  {year} {1996})}\BibitemShut {NoStop}%
\bibitem [{\citenamefont {Boser}\ \emph {et~al.}(1992)\citenamefont {Boser},
  \citenamefont {Guyon},\ and\ \citenamefont {Vapnik}}]{Boser92atraining}%
  \BibitemOpen
  \bibfield  {author} {\bibinfo {author} {\bibfnamefont {B.~E.}\ \bibnamefont
  {Boser}}, \bibinfo {author} {\bibfnamefont {I.~M.}\ \bibnamefont {Guyon}}, \
  and\ \bibinfo {author} {\bibfnamefont {V.~N.}\ \bibnamefont {Vapnik}},\ }in\
  \href@noop {} {\emph {\bibinfo {booktitle} {Proceedings of the 5th Annual ACM
  Workshop on Computational Learning Theory}}}\ (\bibinfo  {publisher} {ACM
  Press},\ \bibinfo {year} {1992})\ pp.\ \bibinfo {pages}
  {144--152}\BibitemShut {NoStop}%
\bibitem [{\citenamefont {Shashua}(2009)}]{Shashua:2009itm}%
  \BibitemOpen
  \bibfield  {author} {\bibinfo {author} {\bibfnamefont {A.}~\bibnamefont
  {Shashua}},\ }\href {http://arxiv.org/abs/0904.3664} {\bibfield  {journal}
  {\bibinfo  {journal} {CoRR}\ }\textbf {\bibinfo {volume} {abs/0904.3664}}
  (\bibinfo {year} {2009})},\ \Eprint {http://arxiv.org/abs/0904.3664}
  {arXiv:0904.3664} \BibitemShut {NoStop}%
\bibitem [{\citenamefont {Platt}(1998)}]{Platt98sequentialminimal}%
  \BibitemOpen
  \bibfield  {author} {\bibinfo {author} {\bibfnamefont {J.~C.}\ \bibnamefont
  {Platt}},\ }\href@noop {} {\  (\bibinfo {year} {1998})}\BibitemShut {NoStop}%
\end{thebibliography}%
\end{document}